\documentclass[aip,reprint,amsmath,amssymb]{revtex4-1}
\usepackage{graphicx}

\newcommand{\V}[1]{\mathbf{#1}} 
\newcommand{\T}[1]{\texttt{#1}} 
\newcommand\Alfven{Alfv\'en }
\newcommand\Alfvenic{Alfv\'enic }
\newcommand{\figref}[1]{Figure~\ref{#1}}
\newcommand{\secref}[1]{\S\ref{#1}}
\renewcommand{\eqref}[1]{equation~(\ref{#1})}

\begin{document}


\title{A Weakened Cascade Model for Turbulence in Astrophysical Plasmas} 



\author{G.~G. Howes}
\email[]{gregory-howes@uiowa.edu}
\affiliation{Department of Physics and Astronomy, University of Iowa, Iowa City, 
Iowa 52242, USA.}
\affiliation{Isaac Newton Institute for Mathematical Sciences, Cambridge, CB3 0EH, U.K.}

\author{J.~M. TenBarge}
\affiliation{Department of Physics and Astronomy, University of Iowa, Iowa City, 
Iowa 52242, USA.}

\author{W.~Dorland}
\affiliation{Department of Physics, University of Maryland, College Park, 
Maryland 20742-3511, USA.}
\affiliation{Isaac Newton Institute for Mathematical Sciences, Cambridge, CB3 0EH, U.K.}


\date{\today}

\begin{abstract}
A refined cascade model for kinetic turbulence in weakly collisional
astrophysical plasmas is presented that includes both the transition
between weak and strong turbulence and the effect of nonlocal
interactions on the nonlinear transfer of energy. The model describes
the transition between weak and strong MHD turbulence and the
complementary transition from strong kinetic \Alfven wave (KAW)
turbulence to weak dissipating KAW turbulence, a new regime of weak
turbulence in which the effects of shearing by large scale motions and
kinetic dissipation play an important role. The inclusion of the
effect of nonlocal motions on the nonlinear energy cascade rate in the
dissipation range, specifically the shearing by large-scale motions,
is proposed to explain the nearly power-law energy spectra observed in
the dissipation range of both kinetic numerical simulations and solar
wind observations.
\end{abstract}


\maketitle 


\section{Introduction}
Plasma turbulence plays an important role in the transfer of energy in
a wide range of space and astrophysical environments, from clusters of
galaxies, to accretion disks around black holes, to the magnetized
corona of our own sun, to the solar wind that fills the
heliosphere. In all of these diverse environments, turbulence is
responsible for the transport of energy from the large scales at which
the turbulence is driven down to the small scales at which dissipation
mechanisms can effectively dissipate the turbulent fluctuations and
lead to heating of the plasma species.  The understanding of plasma
turbulence, its dissipation, and the resulting plasma heating is
therefore a key goal for the space physics and astrophysics communities.

Although the study of the \emph{inertial range} of plasma
turbulence---the range of scales over which the effects of driving and
dissipation are negligible---dates back more than four decades
\cite{Coleman:1968}, study of the \emph{dissipation range} has only 
recently become a focus of the astrophysics and heliospheric physics
communities. One of the major challenges in the investigation of the
dissipation mechanisms at work in space and astrophysical plasmas is
the fact that, at the small scales on which these mechanisms operate,
the dynamics is typically weakly collisional. Under these conditions,
a kinetic description of the plasma dynamics is necessary
\cite{Marsch:1991,Marsch:2006,Howes:2008c,Howes:2008b,Howes:2008d,Schekochihin:2009},
a much more complicated description than the standard fluid
descriptions, such as magnetohydrodynamics (MHD), commonly used in the
study of the inertial range dynamics.

The increasingly vigorous activity studying the dissipation of plasma
turbulence has been driven by several new developments: the
development of new analytical models of kinetic turbulence
\cite{Howes:2008c,Howes:2008b,Schekochihin:2009}, advances in 
high-performance computation that have made possible kinetic
simulations of weakly collisional turbulence
\cite{Howes:2008a,Howes:2011a}, and, most recently,  new
observational studies of turbulence in the solar wind with sufficient
time resolution to extend the satellite-frame frequency power spectra
of magnetic and electric field fluctuations up to the Doppler-shifted
frequencies associated with the electron Larmor radius
\cite{Sahraoui:2009,Kiyani:2009,Alexandrova:2009,Chen:2010,Sahraoui:2010b}.

This paper presents the refinement of a turbulent cascade model for
weakly collisional plasmas, originally devised by Howes \emph{et al.}
2008 \cite{Howes:2008b}. At the small scales where the turbulent
motions are dissipated, two new physical effects may play a
significant role: the transition between weak and strong turbulence,
and the nonlocal contribution to the nonlinear energy transfer rate.
The aim of this paper is to describe how these additional physical
effects are incorporated into a refined model, denoted the
\emph{weakened cascade model}. Energy spectra predicted by the new model
are compared in detail to the numerical energy spectra from nonlinear
gyrokinetic simulations of turbulence in the dissipation range, and
the physical implications of the nonlocal contribution to the
nonlinear energy transfer in the dissipation range are discussed in
detail.

\section{Motivation for a Refined Cascade Model}
A new and exciting challenge in the study of turbulence in the solar
wind is to understand the recent observations of nearly power-law
magnetic energy spectra in the dissipation range \cite{Note0} up to scales
corresponding to the electron Larmor radius 
\cite{Sahraoui:2009,Kiyani:2009,Alexandrova:2009,Chen:2010,Sahraoui:2010b}.
The physics describing the turbulent dynamics of the inertial range is
expected to be self-similar, dominated by the conditions that are
``local'' in scale \cite{Kolmogorov:1941}.  In the dissipation range,
on the other hand, dissipative mechanisms necessarily play a
non-negligible role, and so the usual assumption that the turbulent
energy transfer is dominated by local interactions must be called into
question.

In anisotropic \Alfvenic turbulence, nonlinear interactions typically
strengthen as the cascade progresses to smaller scale until they reach
and maintain a state of critically balanced, strong turbulence. But as
turbulent fluctuation amplitudes are diminished by some physical
dissipation mechanism, one must consider the possibility that the
turbulence may eventually transition back to a state of weak
turbulence. In addition to this transition from strong to weak
turbulence, one must also account for the effect on the nonlinear
energy transfer rate at small scales (where the local turbulent
fluctuation amplitudes are diminished by some dissipation mechanism)
by the undamped, nonlocal turbulent motions at larger scales. In this
section, we briefly review the relevant concepts from modern theories
of anisotropic magnetized turbulence that provide the theoretical
foundation for these arguments. Then we identify the shortcomings of
the original cascade model \cite{Howes:2008b} and outline the
necessary refinements of the model to account for the transition to
weak turbulence at dissipative scales and for the effect of nonlocal motions
on the energy cascade rate.

\subsection{Review of Anisotropic Plasma Turbulence Theories}

Early theories of magnetized plasma turbulence proposed an isotropic
cascade of turbulent energy from low wavenumbers (large scales) to
large wavenumbers (small scales)
\cite{Iroshnikov:1963,Kraichnan:1965}. Measurements of magnetized 
plasma turbulence both in the laboratory
\cite{Robinson:1971,Zweben:1979,Montgomery:1981} and in the solar wind
\cite{Belcher:1971}, as well as early numerical simulations \cite{Shebalin:1983},
demonstrated the existence of significant anisotropy with respect to
the direction of the local mean magnetic field. In the context of
incompressible MHD, the inclusion of anisotropy in the direction of
the nonlinear turbulent energy transfer through wavevector space lead
to a fundamental distinction between the properties of \emph{weak} MHD
turbulence and \emph{strong} MHD turbulence
\cite{Sridhar:1994,Goldreich:1995}.  As the shift in behavior between 
strong and weak turbulence plays a role when the turbulent cascade
suffers dissipation, a brief review of the relevant properties is
provided here.

Although the theory for weak turbulence in incompressible MHD plasmas
was a focal point of controversy in the late 1990s \cite{Sridhar:1994,
Montgomery:1995,Ng:1996,Goldreich:1997,Ng:1997,Galtier:2000,Lithwick:2003},
a refined theory of weak MHD turbulence has emerged. This theory is
based on a perturbative treatment of weak, resonant three-wave
interactions between \Alfven waves propagating in opposite directions
along the local mean magnetic field.  The need to satisfy both the
resonance conditions for frequency and wavevector \cite{Shebalin:1983}
and the requirement that only counter-propagating \Alfven waves
interact nonlinearly \cite{Iroshnikov:1963,Kraichnan:1965} leads to
the prediction that there is no parallel cascade of energy in
wavenumber space (where the directions parallel and perpendicular are
defined with respect to the \emph{local} mean magnetic
field). Therefore, energy is transferred via the turbulent cascade
anisotropically in wavevector space, only to higher perpendicular
wavenumbers $k_\perp$, while the characteristic parallel wavenumber
$k_\parallel$ remains constant. The one-dimensional magnetic energy
spectrum $E_B(k_\perp)$, defined such that the total magnetic energy
$E_B=\int dk_\perp E_B(k_\perp)$, is predicted to scale as
$E_B(k_\perp) \propto k_\perp^{-2}$.  Recent reduced MHD numerical
simulations have confirmed this predicted scaling in the regime of weak
turbulence \cite{Perez:2008}. The nonlinear interactions
strengthen as the perpendicular wavenumber increases, so as the
turbulent cascade progresses to higher $k_\perp$, the dynamics
eventually violate the assumption of weak nonlinear interactions
required for the application of perturbation theory. Therefore, the
theory of weak MHD turbulence predicts that, if a sufficiently large
inertial range exists, weak MHD turbulence will eventually
transition to a state of strong MHD turbulence \cite{Sridhar:1994,Goldreich:1995}.

When the turbulence becomes sufficiently strong, the perturbative
expansion of the nonlinear term is no longer dominated by the
three-wave interactions, and terms of all orders contribute.  This
leads to a broadening of the resonance, relaxing the strict
constraints on the frequency and wavevector, making possible the
cascade of energy in the parallel direction of wavevector space
\cite{Goldreich:1995}. The theory of strong turbulence in
incompressible MHD plasmas is based on the conjecture of a \emph{critical
balance} \cite{Higdon:1984a,Goldreich:1995} between the linear
timescale for \Alfven waves and the nonlinear timescale of turbulent
energy transfer.  The one-dimensional magnetic energy spectrum is
predicted to scale as $E_B(k_\perp) \propto k_\perp^{-5/3}$. The
conjecture of critical balance predicts the development of a
scale-dependent anisotropy given by $k_\parallel \propto
k_\perp^{2/3}$; therefore, the one-dimensional parallel magnetic
energy spectrum, defined by $E_B=\int dk_\parallel E_B(k_\parallel)$,
is predicted to scale as $E_B(k_\parallel) \propto k_\parallel^{-2}$.
The predictions of the scale-dependent anisotropy are supported by
early numerical simulations \cite{Cho:2000,Maron:2001} and by 
more recent studies of the scaling of the the parallel spectrum in the
solar wind \cite{Horbury:2008,Podesta:2009a}. 

The lack of a parallel cascade in weak MHD turbulence and the scale
dependent anisotropy in strong MHD turbulence have an important
consequence: even when turbulence is driven isotropically at low
wavenumber (large scale) with $k_0=k_{\perp 0}= k_{\parallel 0}$, for
a sufficiently large inertial range (typical of most space and
astrophysical plasmas of interest), the turbulent fluctuations at high
wavenumber (small scale) become significantly anisotropic with
$k_\parallel \ll k_\perp$. When the turbulent cascade reaches the
perpendicular scale of the ion Larmor radius $k_\perp \rho_i \sim 1$,
such turbulent fluctuations transition to a cascade of strong kinetic \Alfven
wave (KAW) turbulence, as has been predicted theoretically
\cite{Gruzinov:1998,Quataert:1999,Howes:2008b,Schekochihin:2009} and  
verified with nonlinear kinetic simulations
\cite{Howes:2008a}. Assuming again a critical balance between the
linear and nonlinear timescales, in the absence of dissipation, the
kinetic \Alfven wave cascade is predicted to have a one-dimensional
magnetic energy spectrum that scales as $E_B(k_\perp) \propto
k_\perp^{-7/3}$ and a scale-dependent anisotropy $k_\parallel \propto
k_\perp^{1/3}$ for the nonlinear transfer of energy in wavevector
space \cite{Biskamp:1999,Cho:2004,Krishan:2004,Dastgeer:2005,
Howes:2008b,Schekochihin:2009}. When collisionless dissipation via the
Landau resonance is included, the original cascade model
\cite{Howes:2008b} predicts that, if magnetometer noise floor 
is taken into account, the spectral index of the measured
one-dimensional magnetic energy spectrum could vary from a value of
$-7/3$ for weak dissipation up to approximately $-4$ for strong
dissipation, with the strength of the dissipation depending on the
plasma parameters $\beta_i$ and $T_i/T_e$.

\begin{figure}
\resizebox{3.3in}{!}{\includegraphics*[0.29in,2.in][8.0in,7.2in]{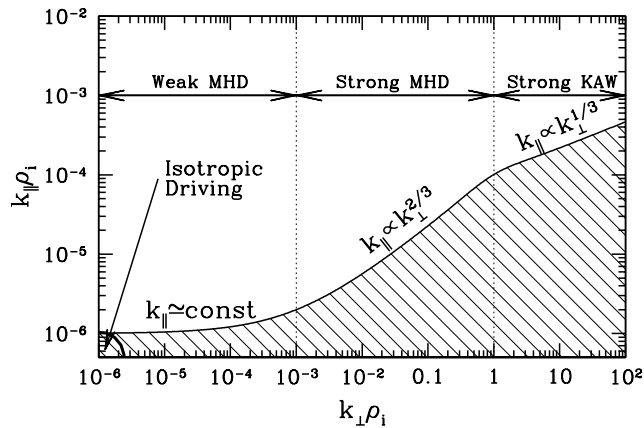}}
\caption{ \label{fig:kspace}  Schematic diagram of the anisotropic transfer 
of energy through wavevector space $(k_\perp,k_\parallel)$. For
turbulence driven isotropically at low wavenumber $k_{\perp 0} \rho_i
\sim k_{\parallel 0} \rho_i \sim 10^{-6}$ (quarter circle in the lower left
of the figure), nonlinear interactions lead to transfer of energy to
higher wavenumbers.  For weak MHD turbulence, the transfer is
restricted to higher $k_\perp$, with no cascade of energy to higher
parallel wavenumbers $k_{\parallel}$.  As the MHD turbulence becomes
strong (at $k_\perp \rho_i \sim 10^{-3}$), resonance broadening allows
a parallel cascade of energy that achieves a state of critical balance
at all scales, with $k_\parallel \propto k_\perp^{2/3}$. Finally, as
the strong MHD turbulence transitions to strong KAW turbulence (at
$k_\perp \rho_i \sim 1$), the turbulence remains in critical balance
but the scaling in this regime changes to $k_\parallel \propto
k_\perp^{1/3}$. }
\end{figure}

To highlight the different behaviors predicted for weak MHD, strong
MHD, and strong KAW turbulence, a schematic of the energy transfer
through wavevector space is presented in \figref{fig:kspace}. Since
the dynamics of an MHD plasma is axisymmetric about the direction of
the local mean magnetic field, the description of energy transfer
through the three-dimensional wavevector space can be reduced to the
two-dimensional plane of the parallel component of the wavevector
$k_\parallel$ (the axial coordinate parallel to the local mean
magnetic field) and the perpendicular component $k_\perp$ (the radial
coordinate perpendicular to the local mean magnetic field). The
anisotropic MHD turbulence theories above suggest that the energy
flows through wavenumber space along the path given by the solid line
in \figref{fig:kspace}---along constant $k_\parallel$ for weak MHD
turbulence and along a path given by critical balance for strong MHD
and strong KAW turbulence. Numerical simulations of MHD turbulence
\cite{Cho:2000,Maron:2001,Cho:2002,Cho:2003,Oughton:2004}, however, 
suggest that the turbulent energy does not flow strictly along this
one-dimensional path through wavevector space but approximately fills
the shaded region in \figref{fig:kspace}. Similar to the original
cascade model, the one-dimensional weakened cascade model presented in
this paper determines the turbulent energy integrated vertically in this
figure over all possible values of $k_\parallel$ at each $k_\perp$;
the effective value of $k_\parallel$ for the integrated turbulent
energy at each $k_\perp$ is assumed to be given by the solid line in
\figref{fig:kspace}.

Several final comments regarding plasma turbulence theories are in
order.  First, note that the plot of the $(k_\perp,k_\parallel)$ plane
in \figref{fig:kspace} is logarithmic on both axes.  The inherently
anisotropic transfer of energy in magnetized plasma turbulence
naturally leads to a condition in which $k_\parallel \ll k_\perp$ at
high wavenumbers.  A number of studies of solar wind turbulence in the
past have assumed a constant angle between the direction of the
wavevector and the local mean magnetic field, which would correspond
to a line with $k_\parallel \propto k_\perp$ on the logarithmic plot
of wavevector space; this leads to a significant underestimate of the
anisotropy at small scales (multi-spacecraft measurements in the solar
wind demonstrate significant anisotropy at the scale of the ion Larmor
radius \cite{Sahraoui:2010b}).  Second, although much of the weak and strong
MHD turbulence theory applies rigorously only to incompressible MHD
plasmas, it is our view that the fundamental concepts derived for
incompressible MHD plasmas offer useful guidance for the understanding
of turbulence in plasmas allowing a wider range of physical effects,
\emph{e.g.},~compressibility, finite Larmor radius effects, linear
kinetic damping. Finally, although we choose the particular strong
turbulence scaling given by Goldreich and Sridhar
\cite{Goldreich:1995}  as the basis for the original and
weakened cascade models, an alternative theory for the scaling of
strong MHD turbulence has been proposed by Boldyrev
\cite{Boldyrev:2006}; application of the weakened cascade model using the 
Boldyrev scaling will be  discussed in a subsequent paper.

\subsection{The Original Cascade Model}

The aim of the original cascade model \cite{Howes:2008b} is to explain
the observed magnetic energy spectrum in the solar wind using a
minimal number of ingredients, namely, finite Larmor radius effects
and kinetic damping via the Landau resonance. This model employs a
one-dimensional continuity equation for the magnetic energy spectrum
in perpendicular wavenumber space, and is based on three assumptions:
(a) the Kolmogorov hypothesis  that the energy transfer is determined
locally in wavenumber space\cite{Kolmogorov:1941}; (b) that a state of
critical balance exists between the linear and nonlinear timescales at
all wavenumbers in the spectrum
\cite{Higdon:1984a,Goldreich:1995}; and (c) that the linear kinetic
damping rates determine the dissipation of the turbulent fluctuations
even in the presence of the nonlinear cascade. Results of the model
generally demonstrate the qualitative feature that, as the linear
kinetic damping becomes strong, the spectrum begins an exponential
fall off.  This qualitative feature is observed neither in recent
nonlinear kinetic simulations
\cite{Howes:2011a} nor in recent observations of the
dissipation range of the solar wind
\cite{Sahraoui:2009,Kiyani:2009,Alexandrova:2009,Chen:2010,Sahraoui:2010b}.

The failure of the original cascade model to reproduce the correct
qualitative behavior when collisionless damping becomes strong is due
to the assumption of strong turbulence satisfying the critical balance
condition at all scales. It seems relatively clear that, as
collisionless damping reduces the turbulent amplitude at a given
wavenumber to a level below that expected for a dissipationless
cascade, the nonlinear turbulent interactions at that wavenumber must
cease to be strong.  One then expects a transition to a cascade with
weak nonlinear interactions (in the sense of weak turbulence),
although the collisionless damping and shearing by larger scale
motions will also play an important role, so it won't be a standard weak
turbulence picture.

The failure of the strong turbulence assumption at perpendicular
wavenumbers where the dissipation becomes significant is made clear by
a simple physical argument applied to the original cascade model
results. In Figure~7 of the original cascade model paper
\cite{Howes:2008b}, the parallel wavenumber as a function of the 
perpendicular wavenumber, $k_\parallel (k_\perp)$, is plotted for
plasma parameters $\beta_i=1$ and $T_i/T_e=1$.  At a value of $k_\perp
\rho_i \simeq 8$, the value of $k_\parallel$ peaks, and then it begins
to drop.   Physically, this would mean that, as the turbulence
continues to cascade to smaller perpendicular scales, its parallel
scale length actually increases. This predicted behavior does not make
physical sense. This point is discussed at length in paragraph [46] of
that paper, with the forward looking conclusion that ``nonlinear
simulations are necessary to determine accurately the behavior of the
turbulent cascade as the kinetic damping becomes significant.''

\subsection{Weak Turbulence and Nonlocal Interactions}
\label{sec:weaknl}

The more likely physical result, when the turbulence is dissipatively
weakened, is that the cascade to higher parallel wavenumber is
suppressed, so the parallel wavenumber remains constant.  Weak
turbulence theory predicts that the parallel cascade is suppressed in
incompressible MHD plasmas \cite{Galtier:2000}, but whether this also
holds in the kinetic \Alfven wave regime is not known. We conjecture
here that the parallel cascade is suppressed for weak kinetic
\Alfven wave turbulence, as discussed in more detail in 
\secref{sec:weak}. In addition to this effect, the
nonlinear transfer of energy in weak turbulence also requires many
uncorrelated ``collisions'' between counter-propagating \Alfven wave
packets, in contrast to the single collision required in strong
turbulence \cite{Sridhar:1994,Goldreich:1995}.

Therefore, the weakened cascade model proposed here describes the
transition between strong and weak turbulence by incorporating two
changes: (1) altering the cascade in parallel wavenumber; and (2)
increasing the number of \Alfven wave collisions required to
accomplish nonlinear energy transfer.  The consequence of these
changes, however, is that the collisionless damping at a given scale
becomes relatively stronger than the nonlinear transfer, causing the
spectra to cut-off even more abruptly, resulting in even greater
discrepancy between the cascade model predictions and both the
numerical simulations and the solar wind observations.

To resolve this discrepancy requires the incorporation of another
physical effect, in addition to the transition to weak turbulence.  It
requires accounting for the effect on the nonlinear energy transfer
rate at a given scale by turbulent motions at other scales. Consider
specifically the nonlinear turbulent energy transfer at a particular
perpendicular wavenumber $k_{\perp *}$.  Although the turbulent energy
is still transferred locally in scale space---for example, energy
transfer from wavenumber $k_{\perp *}$ to $2k_{\perp *}$--- this
nonlinear energy transfer is not only due to motions at the local
wavenumber $k_{\perp *}$, but also due to nonlocal (in scale space)
motions at smaller and larger wavenumbers.  Both coherent shearing by
motions at low wavenumbers $k_\perp < k_{\perp *}$ (large scales) and
incoherent diffusion by motions at high wavenumbers $k_\perp >
k_{\perp *}$(small scales) may contribute substantially to the total
nonlinear energy transfer rate from $k_{\perp *}$ to $2k_{\perp *}$.
Therefore, we must abandon the Kolmogorov hypothesis of locality and
account for the effect of nonlocal interactions on the turbulent
energy transfer.

The physical effect due to shearing by large scale motions is
essentially the same as that seen in high magnetic Prandtl number
dynamo simulations using MHD \cite{Schekochihin:2004d}. The magnetic
Prandtl number is defined by the ratio of viscosity over the magnetic
diffusivity, $\mbox{Pr}_m=\nu/\eta$, and typically $\mbox{Pr}_m\gg 1$
for many astrophysical plasmas of interest. For a high magnetic
Prandtl number plasma, magnetic energy can be supported on subviscous
scales, and this magnetic energy is observed in simulations to cascade
to ever smaller scales, until the resistive scale is reached, at which
point the magnetic energy can be dissipated. This cascade occurs
although the plasma cannot support any fluid motions at the subviscous
scales of interest.  The transfer of magnetic energy at these small,
subviscous scales is accomplished by the shearing due to larger scale
motions (at scales larger than the viscous cut-off).  This energy
transfer is therefore nonlocal in nature.

Such an effect must come into play in a weakly collisional plasma when
the turbulent cascade reaches a small enough perpendicular scale that
collisionless damping can diminish the amplitude of the local (in
scale) turbulent fluctuations. Kolmogorov's locality assumption is
well supported in the inertial range, because the energy cascade rate
due to shearing by local fluctuations always dominates over the energy
cascade rate due to shearing by larger-scale motions.  But, in the
dissipation range, where the local fluctuations begin to diminish in
amplitude due to some dissipative mechanism, their dominance of the
local energy cascade rate breaks down, and the effects of shearing by
larger-scale, undamped motions must be taken into account. In this
way, the energy cascade rate at the small, rather strongly damped
scales may be dominated by this nonlocal shearing, with the weak
turbulent interactions due to local fluctuations playing a subdominant
role. Thus, the energy can be cascaded to ever smaller scale at a
reasonably large rate even though the nonlinear energy transfer rate
due to the local fluctuations becomes negligible. This nonlocal
shearing effect is proposed to explain the nearly power-law appearance
of the numerical and observational spectra, and is critical for
understanding the turbulence in the dissipation range.  The nonlocal contribution
to the energy cascade rate by diffusive motions at small scales,
although negligible at the far end of the dissipation range, is
also included in the refined model for consistency.

\section{The Weakened Cascade Model}
In order to refine the original cascade model \cite{Howes:2008b}, we
must incorporate two physical effects not included in the original model:
\begin{enumerate}
\item Weak Turbulence: The model must be able to handle the quantitative
changes in the energy cascade rate in both the weak and strong
turbulence regimes.  Of particular importance is the qualitative
difference in the parallel cascade of energy: in weak turbulence, there
is no cascade of energy to smaller parallel scales; and in strong
turbulence, the parallel cascade is governed by critical balance.
\item Nonlocal Interactions: The net energy cascade rate at a given 
wavenumber $\epsilon(k_{\perp *})$ must account for the nonlinear
transfer due to both the local fluctuations at $k_{\perp *}$ and
nonlocal fluctuations at other wavenumbers. The nonlocal fluctuations
contribute to the nonlinear energy cascade rate due to shearing by
fluctuations at smaller wavenumbers $k_{\perp } < k_{\perp *}$ and
diffusion by fluctuations at larger wavenumbers $k_{\perp } > k_{\perp
*}$.

\end{enumerate}
In this section, we describe in detail the quantitative modifications of
the original cascade model required to incorporate these two physical
effects, resulting in the new \emph{weakened cascade model}.

\subsection{Weak Turbulence}
\label{sec:weak}
The key parameter in distinguishing weak from strong turbulence is
the nonlinearity parameter
\begin{equation}
\chi \sim \frac{k_\perp v_\perp}{\omega}.
\end{equation}
This dimensionless parameter $\chi$ measures the ratio of the
nonlinear frequency $\omega_{nl} \simeq k_\perp v_\perp$ to the linear
wave frequency $\omega$. Strong turbulence corresponds to $\chi \sim
1$, satisfying the condition of critical balance $\omega \sim
\omega_{nl}$, whereas weak turbulence corresponds to $\chi \ll
1$. Note that the case of overdriven turbulence $\chi > 1$ has not been
thoroughly explored or discussed in the literature; it probably
deserves some attention, but henceforth we will consider only the
cases of weak or critically balanced turbulence $\chi \lesssim 1$.

To handle the transition to kinetic \Alfven wave turbulence, we focus
on the magnetic energy rather than the kinetic energy, so we want to
write the nonlinear energy cascade rate in terms of the magnetic
fluctuation energy, $b_k^2 \equiv \delta B^2_\perp(k_\perp)/4 \pi n_i
m_i$, where we have written the magnetic field fluctuation in velocity
units. We also adopt the shorthand $v_k \equiv v_\perp
(k_\perp)$. Following the original cascade model, we relate the
velocity and magnetic field fluctuations to each other using the
linear theory,
\begin{equation} 
v_k = \pm \alpha(k_\perp) b_k,
\end{equation}
where the coefficient $\alpha$ smoothly transitions from the MHD to
the KAW limit,
\begin{equation}
\alpha(k_\perp)   = \left\{ 
\begin{array}{cc} 
1,& k_\perp\rho_i\ll 1  \\
k_\perp \rho_i/
{\sqrt{\beta_i+ 2/(1 + T_e/T_i)}}, & k_\perp\rho_i\gg 1
\end{array} \right.
\end{equation}
Note that the linear frequency in the gyrokinetic limit \cite{Note1}
is given by
\begin{equation}
\omega = \pm \overline{\omega}(k_\perp) k_\parallel v_A .
\end{equation}
where $\overline{\omega}(k_\perp) =\alpha(k_\perp)$ in both asymptotic
ranges $k_\perp\rho_i\ll1$ and $k_\perp\rho_i\gg1$ but not in the
transition region $k_\perp\rho_i\sim 1$.  For this simple model, we
take the approximation that $\alpha (k_\perp) =
\overline{\omega}(k_\perp)$ over all scales.

Thus, an appropriate definition of the nonlinearity parameter valid
for all scales is
\begin{equation}
\chi = \frac{C_2 k_\perp v_k}{\omega} =\frac{C_2 k_\perp b_k \alpha}{ k_\parallel v_A  \alpha}
=\frac{C_2 k_\perp b_k}{ k_\parallel v_A},
\end{equation}
where $C_2$ is an order-unity, dimensionless ``Kolmogorov'' constant.

There are two primary effects that must be captured in order to incorporate 
the weak turbulence limit into the cascade model:
\begin{enumerate}
\item Slower nonlinear energy cascade rate due to the effect of many
uncorrelated weak interactions between oppositely directed \Alfven waves
\cite{Sridhar:1994}.
\item The suppression of the energy cascade to small parallel scales.
\end{enumerate}
These issues are addressed separately below.

We define the energy cascade rate by 
\begin{equation}
\epsilon(k_\perp) = C_1^{-3/2} \omega_{nl} b_k^2.
\label{eq:wc_eps}
\end{equation}
where $C_1$ is an order-unity, dimensionless Kolmogorov constant and
the nonlinear frequency, bridging weak and strong turbulence, is given
by
\begin{equation}
\omega_{nl} = \chi k_\perp  v_k = \left( \frac{C_2 k_\perp b_k}{ k_\parallel v_A} \right)
k_\perp b_k \alpha.
\end{equation} 
Note that the cascade time is given by 
\begin{eqnarray}
\tau_{nl} &= & \omega_{nl}^{-1}  =\left( \frac{ k_\parallel v_A} {C_2  k_\perp b_k} \right)
(k_\perp b_k \alpha )^{-1} \\
& = &\left( \frac{ k_\parallel v_A} {C_2 k_\perp b_k} \right)^2 
C_2 (k_\parallel v_A  \alpha )^{-1} = N^2 C_2 (k_\parallel v_A \alpha  )^{-1}.
\end{eqnarray}
This shows explicitly that, in the MHD limit $\alpha=1$, it takes
$N^2 = 1/\chi^2$ Alfven wave packet collisions, each lasting an \Alfven wave
crossing time $\tau_A=(k_\parallel v_A)^{-1}$, for the energy at a
given scale to be transferred to the next scale.

In summary, the energy cascade rate may  be written 
\begin{equation}
\epsilon(k_\perp) = C_1^{-3/2} \left( \frac{C_2 k_\perp b_k}{ k_\parallel v_A} \right)
k_\perp b_k^3 \alpha.
\end{equation}
Let us investigate the predicted steady-state magnetic energy spectrum
in different limits for a constant energy cascade rate $\epsilon$.

In the MHD limit, $\alpha =1$.  The critically balanced, strong
turbulence limit gives $\chi = C_2 k_\perp b_k/ k_\parallel v_A = 1$, so
we have $\epsilon = C_1^{-3/2}k_\perp b_k^3$, yielding a solution for
the magnetic field $ b_k = C_1^{1/2} \epsilon^{1/3} k_\perp^{-1/3}$
and a corresponding 1-D magnetic energy spectrum $E_B(k_\perp) =
b_k^2/k_\perp =C_1
\epsilon^{2/3} k_\perp^{-5/3}$. In the weak turbulence limit $\chi <
1$, we have $\epsilon = C_1^{-3/2} C_2 k_\perp^2 b_k^4 / k_\parallel v_A$
and $k_\parallel$ remains constant, so we obtain a solution for the
magnetic field $ b_k = C_1^{3/8} C_2^{-1/4} \epsilon^{1/4} (k_\parallel
v_A)^{1/4} k_\perp^{-1/2}$ and a corresponding 1-D magnetic energy
spectrum $E_B(k_\perp) =C_1^{3/4}C_2^{-1/2}
\epsilon^{1/2} (k_\parallel v_A)^{1/2} k_\perp^{-2}$. 

In the KAW limit, we approximate $\alpha \simeq k_\perp \rho_i$.
Strong, critically balanced KAW turbulence also has $\chi = C_2 k_\perp b_k/
k_\parallel v_A = 1$, so we have $\epsilon = C_1^{-3/2}k_\perp^2
\rho_i b_k^3$, yielding a solution for the magnetic field $ b_k =
C_1^{1/2} \epsilon^{1/3} k_\perp^{-2/3}
\rho_i^{-1/3}$ and a corresponding 1-D magnetic energy spectrum
$E_B(k_\perp) =C_1 \epsilon^{2/3} k_\perp^{-7/3} \rho_i^{-2/3}$.  In
the weak turbulence limit $\chi < 1$, we have $\epsilon = C_1^{-3/2}
C_2 k_\perp^3 b_k^4 \rho_i/ k_\parallel v_A$.  In this case, we
\emph{assume} that $k_\parallel$ remains constant in the weak
turbulence limit; this assumption is discussed later in this
subsection. The solution for the magnetic field is then $ b_k =
C_1^{3/8} C_2^{-1/4} \epsilon^{1/4} (k_\parallel v_A)^{1/4} k_\perp^{-3/4}
\rho_i^{-1/4}$ and a corresponding 1-D magnetic energy spectrum
$E_B(k_\perp) =C_1^{3/4} C_2^{-1/2} \epsilon^{1/2} (k_\parallel v_A)^{1/2 }
k_\perp^{-5/2} \rho_i^{-1/2}$, agreeing with theoretical predictions
for weak KAW turbulence \cite{Galtier:2006}.

The other important effect when synthesizing a theory that combines
the weak and strong turbulence limits and their influence on the
turbulent cascade of energy is to model the parallel cascade of
energy. In the strong limit, the condition of critical balance allows
the determination of $k_\parallel(k_\perp)$, strictly a function of
$k_\perp$.  The assumption of critical balance at \emph{all} scales in
the original cascade model allowed the damping term to be simplified.
The original cascade model equation for the evolution of the magnetic
energy was written as
\begin{equation}
\frac{\partial b_k^2}{\partial t} = 
-k_\perp \frac{\partial \epsilon(k_\perp) }{\partial k_\perp} + S(k_\perp) - 
2{\gamma} b_k^2,
\end{equation}
where the last term could be simplified, under the assumption of critical 
balance and for the linear kinetic damping rate in the gyrokinetic limit $\gamma = \pm
\overline{\gamma}(k_\perp) k_\parallel v_A$,  to take the form
\begin{equation}
 - 2C_1^{3/2}C_2\frac{\overline{\gamma}(k_\perp)}{\overline{\omega}(k_\perp)} 
\epsilon(k_\perp),
\end{equation}
strictly a function of $k_\perp$.  When this assumption is relaxed,
however, to determine the damping rate one needs to know the value of
$k_\parallel(k_\perp)$ to find
\begin{equation}
\gamma(k_\perp, k_\parallel) =\overline{\gamma}(k_\perp)k_\parallel v_A.
\label{eq:damping}
\end{equation}
Without critical balance, we must devise some other means of determining 
an appropriate value of $k_\parallel$.

It must be noted that the distribution of turbulent power in the
plasma can fill a region of the two-dimensional wave vector space
$(k_\perp, k_\parallel)$, as denoted by the shaded region in
\figref{fig:kspace}.  Here we follow the original cascade model in
treating the magnetic fluctuation energy integrated over all possible
values of $k_\parallel$, so that
\begin{equation}
b_k^2 (k_\perp) = \int dk_\parallel \delta B^2_\perp
(k_\perp,k_\parallel)/ 4 \pi n_i m_i.
\end{equation}
We will treat the total magnetic energy as if it resides at a single
value of $k_\parallel$. The particular choice we make is the maximum
value of $k_\parallel$ that contains significant fluctuation energy.
Therefore, in the strong turbulence limit, $k_\parallel$ is chosen so
that the parallel cascade satisfies the critical balance criterion.
In the weak turbulence limit, however, we assume that the parallel
cascade of energy is inhibited, so the value of $k_\parallel$ remains
constant as energy cascades to larger $k_\perp$. By making this
choice, we again can solve for the turbulent cascade of magnetic
energy as a one-dimensional problem in $k_\perp$, but as we shall see
this necessitates solving for $k_\parallel$ from the driving wavenumber on
up.

It has been proven rigorously that there is no parallel cascade of
energy for weak turbulence in the limit of incompressible MHD
\cite{Galtier:2000}. Heuristic arguments for weak turbulence in
incompressible Hall MHD plasmas
\cite{Galtier:2006} and numerical evidence demonstrating an
anisotropic cascade of energy in weak  whistler wave turbulence
\cite{Gary:2008,Saito:2008,Gary:2010} suggest that the parallel
cascade is also suppressed for the dispersive wave modes at
perpendicular scales smaller than the ion Larmor radius. Therefore, we
conjecture that the parallel cascade is suppressed in weak kinetic
\Alfven wave turbulence as well \cite{Note6}. Nonlinear kinetic simulations of
turbulence in the kinetic \Alfven wave regime will play a key role in
testing this hypothesis.

To model these effects, we take the equation for the evolution of the
parallel wavenumber to be
\begin{equation}
\frac{d \ln k_\parallel }{d \ln k_\perp} = 
\left[ \frac{ 2/3 + (1/3)(k_\perp \rho_i)^2}{
1 + (k_\perp \rho_i)^2 }\right] \chi^2
\end{equation}

Let us now consider the limits of this equation. In the limit of
critically balanced, strong turbulence, $\chi =1$.  In the MHD limit
$k_\perp \rho_i \ll 1 $, we obtain $d \ln k_\parallel /d \ln k_\perp =
2/3$, or $k_\parallel \propto k_\perp^{2/3}$; in the KAW limit,
$k_\perp \rho_i \gg 1 $, we find $d \ln k_\parallel /d \ln k_\perp =
1/3$, or $k_\parallel \propto k_\perp^{1/3}$. In the weak turbulence
limit in both regimes, $\chi^2 \rightarrow 0$, so we find the result
that $k_\parallel$ remains constant \cite{Note2}.

To apply this equation to the cascade model, we begin at the driving
scale $(k_{\perp 0}, k_{\parallel 0})$, and integrate forward over
the logarithmically spaced grid points in $k_\perp$
\begin{equation}
 k_{\parallel j+1} =  k_{\parallel j} \left\{ 1 + 
\left[ \frac{ 2/3 + (1/3)(k_{\perp j} \rho_i)^2}{
1 + (k_{\perp j} \rho_i)^2 }\right] \chi_j^2 \Delta \ln k_\perp \right\},
\end{equation}
where the nonlinearity parameter must be calculated at each gridpoint $j$,
\begin{equation}
\chi_j =\frac{C_2 k_{\perp j} b_{k j}}{ k_{\parallel j} v_A}.
\end{equation}
Note that the value of the nonlinearity parameter is constrained to
have values $\chi_j \le 1$ at all points.

Note also that $k_\parallel$ is never allowed to decrease as $k_\perp$
increases. This behavior appears to make good sense physically.  Consider
turbulent fluctuations at a given scale characterized by a
perpendicular scale $l_{\perp *}$ (or $k_{\perp *}$) and a parallel
scale $l_{\parallel *}$ (or $k_{\parallel *}$). For a fluctuation at a
smaller perpendicular scale $l_\perp < l_{\perp *}$ (or $k_\perp >
k_{\perp *}$), it seems to be unphysical that this smaller perpendicular
scale fluctuation could have a larger parallel scale $l_\parallel
>l_{\parallel *}$ (or $k_\parallel < k_{\parallel *}$).

\subsection{Nonlocal Interactions}
The Kolmogorov hypothesis states that the rate of nonlinear energy
transfer at a given perpendicular wavenumber $k_{\perp *}$ depends
only on the conditions at that wavenumber.  This assumption of
locality in scale space leads to the familiar self-similar scaling of
the turbulent cascade within the inertial range. When the effects of
dissipation are taken into account, however, it becomes necessary to
abandon this limiting hypothesis and to account for the effect of
nonlocal motions, at both lower and higher wavenumbers, on
the nonlinear energy cascade rate at the local wavenumber $k_{\perp *}$.

We first note that the energy transfer still occurs locally in scale
space, with energy being transferred from wavenumber $k_{\perp *}$ to
$2k_{\perp *}$. This energy transfer, however, is not mediated solely
by the local turbulent motions at wavenumber $k_{\perp *}$. Rather,
motions from both smaller wavenumbers $k_\perp < k_{\perp *}$ (large
scales) and larger wavenumbers $k_\perp > k_{\perp *}$(small scales)
may contribute substantially to the total nonlinear energy transfer
rate at wavenumber $k_{\perp *}$. Therefore, the cascade described by
the weakened cascade model is essentially nonlocal in character,
formally precluding the possibility of self-similar solutions. Below
we discuss the effect of both larger and smaller scale
motions on the local turbulent energy transfer.

A generic property of the turbulent cascade---whether weak or strong
or in the MHD or KAW regimes---is that the nonlinear timescale
$\tau_{nl} \sim \omega_{nl}^{-1}$ always decreases as the wavenumber
increases. Therefore, the longer characteristic timescale of large
scale turbulent motions (relative to the local scale) means that their
effect on the local turbulent fluctuations will be coherent over the
lifetime of the local fluctuations. The large scale turbulent
fluctuations can effectively be considered as a shearing motion
applied to the local fluctuations. The contribution to the nonlinear
frequency due to these large scale shearing motions can be accounted
for by summing over all larger scale motions in the cascade,
\begin{equation}
\omega_{nl}^{(s)}(k_\perp) =  \int_{k_{\perp 0}}^{k_{\perp {max}}}
d \ln k_\perp'
\omega_{nl}^{(\mbox{loc})}( k_\perp')  \Theta( k_\perp- k_\perp'),
\end{equation}
where $k_{\perp 0}$ is the driving (outer) scale of the turbulence,
$k_{\perp {max}}$ is the smallest (inner) scale, and $ \Theta$ is the
piecewise constant Heaviside step function.  Here, the nonlinear frequency
due to shearing motions at a scale $k_\perp'$ is given by
\begin{equation}
\omega_{nl}^{(\mbox{loc})}( k_\perp')=
\chi( k_\perp')  k_\perp' b_k( k_\perp') \alpha ( k_\perp').
\end{equation}

The effect of smaller scale motions on the local scale, on the other
hand, can be treated as a diffusive process because the characteristic
lifetime of the small scale motions is shorter than the timescale of
the local fluctuations. The diffusion coefficient due to motions at a
scale $l$ with a timescale $\tau_{l}$ is given by $\mathcal{D}\sim
l^2/\tau_l$. When translated into nonlinear frequencies and
wavenumbers, the diffusion coefficient due to motions at a large
wavenumber $ k_\perp'$ (small scale) takes the form
$\mathcal{D}(k_\perp') \sim
\omega_{nl}^{(\mbox{loc})}( k_\perp')/k_\perp'^2$. Summing over
all smaller scales leads to the contribution to the nonlinear
frequency due to these small scale diffusive motions, given by
\begin{equation}
\omega_{nl}^{(d)}(k_\perp)  = \int_{k_{\perp 0}}^{k_{\perp {max}}}
d \ln k_\perp'
\omega_{nl}^{(\mbox{loc})}( k_\perp')\frac{ k_\perp^2 }{k_\perp'^2}  \Theta( k_\perp'- k_\perp).
\end{equation}

Therefore, the total nonlinear frequency may be written as the sum of
terms due to large-scale shearing, local-scale fluctuations, and small-scale diffusion,
\begin{equation} \omega_{nl}(k_\perp)=
\omega_{nl}^{(s)}(k_\perp) + \omega_{nl}^{(l)}(k_\perp) +
\omega_{nl}^{(d)}(k_\perp),
\end{equation}
where the local term may be written in an analogous manner,
\begin{equation}
\omega_{nl}^{(l)}(k_\perp)  =  \int_{k_{\perp 0}}^{k_{\perp {max}}}
d \ln k_\perp'
\omega_{nl}^{(\mbox{loc})}( k_\perp')  \delta( k_\perp- k_\perp').
\label{eq:wnl_loc}
\end{equation}


It is important to note that the inclusion of these nonlocal
interaction terms does not alter the scaling of the turbulent cascade
in the MHD inertial range (or in the KAW inertial range in the
absence of dissipation), as shown in Appendix~\ref{app:nonloc_scaling}.

\subsection{Summary of the Weakened Cascade Model}
\label{sec:wc_summary}
In this section, we summarize the weakened cascade model for clarity
and ease of  reference. The key dependent variables of this
one-dimensional model are the perpendicular magnetic field energy
$b_k^2$ (integrated over all $k_\parallel$) and the parallel
wavenumber $k_\parallel$, both functions of the independent variable,
the perpendicular wavenumber $k_\perp$. The continuity equation for
magnetic energy in perpendicular wavenumber space is given by 
\begin{equation}
\frac{\partial b_k^2}{\partial t} = 
-k_\perp \frac{\partial \epsilon }{\partial k_\perp} + S - 
2{\gamma} b_k^2,
\label{eq:kcont}
\end{equation}
where the steady state magnetic energy spectrum is determined by
iterating numerically until the right-hand side equals zero for all
wavenumbers.  Here the source term $S(k_\perp)$ determines the energy
input at the driving scale, characterized by driving wavenumber
components $k_{\perp 0}$ and $k_{\parallel 0}$. The linear kinetic
damping rate is a function of both $k_\perp$ and $k_\parallel$ and may
be written as
\begin{equation}
\gamma(k_\perp, k_\parallel) =\overline{\gamma}(k_\perp)k_\parallel v_A.
\end{equation}
Note that, in the gyrokinetic limit $k_\parallel \ll k_\perp$, the
normalized damping rate $\overline{\gamma}\equiv \gamma/(k_\parallel
v_A)$ is only a function of $k_\perp$ and may be determined by the
gyrokinetic or Vlasov-Maxwell dispersion relation \cite{Howes:2006,Howes:2008b}.  The energy cascade
rate $\epsilon(k_\perp)$ is a subsidiary function  defined by
\begin{equation}
\epsilon(k_\perp) = C_1^{-3/2} \omega_{nl} b_k^2.
\end{equation}
and the total nonlinear frequency $\omega_{nl} $ at wavenumber $k_\perp$---including
terms due to large-scale shearing, local-scale fluctuations,
and small-scale diffusion---is given by
\begin{eqnarray}
 \omega_{nl}(k_\perp)& =&  \int_{k_{\perp 0}}^{k_{\perp {max}}}
d \ln k_\perp'
\omega_{nl}^{(\mbox{loc})}( k_\perp') \nonumber \\
& &  \times \left[ \Theta( k_\perp- k_\perp') 
+ \frac{ k_\perp^2 }{k_\perp'^2}  \Theta( k_\perp'- k_\perp)
\right] \label{eq:wnl}
\end{eqnarray}
where the piecewise constant Heaviside step function is defined by
\begin{equation}
\Theta( x) = \left\{ 
\begin{array}{ll}
1 & x>0\\
1/2 & x=0 \\
0 & x < 0
\end{array}
\right. ,
\end{equation}
and where the contribution to the nonlinear frequency due to motions
at each wavenumber $k_\perp'$ is given by
\begin{equation}
\omega_{nl}^{(\mbox{loc})}( k_\perp')=
\chi( k_\perp')  k_\perp' b_k( k_\perp') \overline{\omega} ( k_\perp'),
\end{equation}
and the nonlinearity parameter is defined by
\begin{equation}
\chi( k_\perp')= \min \left( 1,\frac{C_2   k_\perp' b_k( k_\perp')}
{k_\parallel(k_\perp') v_A}\right). 
\end{equation}
In the gyrokinetic limit $k_\parallel \ll k_\perp$, the linear \Alfven
wave frequency is given by $ \overline{\omega}(k_\perp)=
\omega/(k_\parallel v_A)$ and may be determined by the gyrokinetic or
Vlasov-Maxwell dispersion relation \cite{Howes:2006,Howes:2008b}.
Finally, the value of $k_\parallel(k_\perp)$ is found by integrating
from the driving scale $(k_{\perp 0},k_{\parallel 0})$ using the
equation
\begin{equation}
\frac{d \ln k_\parallel }{d \ln k_\perp} = 
\left[ \frac{ 2/3 + (1/3)(k_\perp \rho_i)^2}{
1 + (k_\perp \rho_i)^2 }\right] \chi^2.
\label{eq:dkzdkx}
\end{equation}
The model has only two free parameters, the dimensionless order-unity
Kolmogorov constants: $C_1$ adjusts the relative weight of the nonlinear 
energy transfer to the linear kinetic damping, and $C_2$ adjusts the
condition of critical balance.

\section{Numerical Simulations}
\label{sec:numsim}

To test the ability of the weakened cascade model to predict the
properties of the turbulent steady-state in a weakly collisional
plasma, we employ nonlinear gyrokinetic numerical simulations using
the code \T{AstroGK} \cite{Numata:2010}. In particular, we are
interested in the ability of the weakened cascade model to reproduce
the turbulent energy spectra when the dissipation becomes strong, so
we focus on two simulations that each cover the entire dissipation
range of scales from the ion to the electron Larmor radius. For one
simulation we choose parameters $\beta_i =1$ and $T_i/T_e=1$, leading
to moderate collisionless damping of the turbulent cascade; for the
second simulation we choose parameters $\beta_i =0.01$ and
$T_i/T_e=1$, leading to strong collisionless damping. All the details
of the $\beta_i =1$ simulation are outlined in Howes \emph{et al.}
2011 \cite{Howes:2011a}, so here we focus on the description of the
new $\beta_i=0.01$ simulation.

\T{AstroGK} evolves the perturbed gyroaveraged distribution
function $h_s(x,y,z,\lambda,\varepsilon)$ for each species $s$, the
scalar potential $\varphi$, parallel vector potential $A_\parallel$,
and the parallel magnetic field perturbation $\delta B_\parallel$
according to the gyrokinetic equation and the gyroaveraged Maxwell's
equations \cite{Frieman:1982,Howes:2006}. The velocity space
coordinates are $\lambda=v_\perp^2/v^2$ and $\varepsilon=v^2/2$. The
domain is a periodic box of size $L_{\perp }^2 \times
L_{\parallel }$, elongated along the straight, uniform mean magnetic
field $B_0$. Note that, in the gyrokinetic formalism, all quantities may
be rescaled to any parallel dimension satisfying $L_{\parallel }
/L_{\perp } \gg 1$. Uniform Maxwellian equilibria for ions (protons)
and electrons are chosen, and the correct mass ratio $m_i/m_e=1836$ is
used. Spatial dimensions $(x,y)$ perpendicular to the mean field are
treated pseudospectrally; an upwinded finite-difference scheme is used
in the parallel direction, $z$. Collisions are incorporated using a
fully conservative, linearized collision operator that includes
energy diffusion and pitch-angle scattering
\cite{Abel:2008,Barnes:2009}. 

Both simulations employ a simulation domain size $L_{\perp}=2\pi
\rho_i$ with dimensions $(n_x,n_y,n_z,n_\lambda,n_\varepsilon,n_s)=
(128,128,128,64,16,2)$. The fully dealiased range yields perpendicular
wavenumbers $1 \le k_x \rho_i \le 42$ and $1 \le k_y \rho_i \le 42$.
For a mass ratio of $m_i/m_e=1836$ and $T_i/T_e=1$, the scale of the
electron Larmor radius $k_\perp \rho_e=1$ corresponds to a value of
$k_\perp \rho_i \simeq 42.8$, so the simulation covers the entire
dissipation range of scales from the ion to the electron Larmor
radius.  The importance of covering this entire range is that
wave-particle interactions via the Landau resonance are resolved and
sufficient to damp the electromagnetic fluctuations within the
simulated range of scales.  Therefore, all dissipation in the
simulation is accomplished by resolved Landau damping, eliminating the
need for an \emph{ad hoc} fluid model of the damping, such as
viscosity or resistivity \cite{Howes:2008d}.

Both simulations are driven at only the largest
perpendicular scale in the domain, $k_\perp \rho_i=1$, using six modes
of a parallel ``antenna'' current $j_{\parallel,\mathbf{k}}^a$ added
via Amp\`ere's Law \cite{Numata:2010}.  For the $\beta_i=0.01$  simulation, these driven modes have
wavevectors $(k_x\rho_i,k_y \rho_i,k_z L_\parallel/2 \pi)=(1,0,\pm 1), (0,1,\pm 1),
(-1,0,\pm 1)$, frequencies $\omega_a = 1.2 \omega_{A0}$ (where
$\omega_{A0} \equiv k_{\parallel 0}v_A$ is a characteristic \Alfven
frequency corresponding to the parallel size $L_\parallel$ of the
domain), and amplitudes that evolve according to a Langevin
equation. This produces Alfv\'enic wave modes with a frequency $\omega
\sim \pm k_{\parallel 0} v_A$ and a decorrelation rate comparable to
$\omega$, as expected for critically balanced Alfv\'enic turbulence
\cite{Goldreich:1995}.

The coefficients for the collision operator
\cite{Abel:2008,Barnes:2009} in the $\beta_i=0.01$  simulation are $\nu_i=0.01\omega_{A0}$ and
$\nu_e=0.1 \omega_{A0}$, chosen to achieve sufficient damping of
small-scale velocity-space structure yet to avoid altering the
collisionless dynamics of each species over the range of scales at
which the kinetic damping is non-negligible.

Both simulations are brought to a statistically steady state with minimal
computational expense by using a recursive expansion procedure
\cite{Howes:2008c}. At low spatial resolution, the simulation
is run for more than an outer-scale eddy turnover time (this turnover
time is $\tau_0=4.65 \omega_{A0}^{-1}$ for the $\beta_i=0.01$
simulation) to reach a steady state. Resolution in each spatial
dimension is then doubled, and the simulation is run to a new steady
state, which requires only a time of order the cascade time at the
smallest resolved mode before expansion. Three applications of this
expansion procedure brings the perpendicular dynamic range from an
initial value of 6 to a final value of 42.  The $\beta_i=0.01$
simulation has been evolved using this procedure to a time of
$t=166.23 \omega_{A0}^{-1}$. 

The resulting magnetic and electric energy spectra from both
simulations are presented in the next section (Figures~\ref{fig:milestone} and 
\ref{fig:sw42sld}). The normalized one-dimensional magnetic-energy spectrum is defined by
$E_{B_\perp}(k_\perp) = (L_\parallel/L_\perp^2) 2\pi (k_\perp \rho_i)^3\int
dz\, {\langle|A_{\parallel,\V{k}_\perp}(z)|^2\rangle/8\pi n_i T_i}$,
where the angle brackets denote a sum of the energy of all
perpendicular Fourier modes falling in a wavenumber shell centered at
$|\V{k}_\perp|=k_\perp$ with width $2\pi/L_\perp$.  The normalized
electric-energy spectrum $E_{E_\perp}(k_\perp)$ is defined similarly
in terms of $\varphi_{\V{k}_\perp}$, with an extra factor of
$(c/v_A)^2$, where $c$ is the speed of light. Note that since the
Fourier modes are defined in Cartesian coordinates, but the value of
$k_\perp$ is polar, there are modes in the ``corner'' of the
simulation (in Fourier space) where full information at all azimuthal
angles is not available---therefore beyond the value of $k_\perp
\rho_i = 42$, not all modes are represented, leading to a drop in the
energy in the 1-D energy spectrum. We choose to plot the spectra over
this range to demonstrate that no bottleneck of electromagnetic
fluctuation energy at small scales occurs in these simulations.

\section{Results of Weakened Cascade Model}
\label{sec:results}
In this section we present results of the weakened cascade model, as
defined by equations~(\ref{eq:kcont})--(\ref{eq:dkzdkx}), for cases of
interest.  All weakened cascade model calculations presented in this
section employ the linear collisionless gyrokinetic dispersion
relation \cite{Howes:2006} to calculate the normalized linear kinetic
frequency $\overline{\omega}=\omega/(k_\parallel v_A)$ and damping
rate $\overline{\gamma}=\gamma/(k_\parallel v_A)$ as a function of
three dimensionless plasma parameters: the normalized perpendicular
wavenumber $k_\perp \rho_i$, the ion plasma beta $\beta_i$, and the
ion to electron temperature ratio $T_i/T_e$.  A fully ionized plasma
of protons and electrons with isotropic Maxellian equilibrium velocity
distributions is assumed, and a realistic mass ratio of $m_i/m_e=1836$
is used.

\subsection{Transition from Weak to Strong MHD Turbulence}
The first important test of the weakened cascade model is to verify
that it reproduces the theoretically predicted characteristics of the
transition from weak MHD turbulence to strong MHD turbulence. The plasma
parameters chosen for this test are $\beta_i=1$ and $T_i/T_e=1$, and
the turbulence is driven isotropically at $k_{\perp 0} \rho_i=
k_{\parallel 0} \rho_i= 10^{-6}$. The amplitude of the source term is
chosen so that the nonlinearity parameter at the driving scale is
$\chi(k_{\perp 0})=0.1$, giving rise to a weak MHD turbulent cascade.
The Kolmogorov constants are taken to be $C_1=1.4$ and $C_2=1$, and
the spectrum is solved using 60 logarithmically spaced gridpoints over
$k_\perp\rho_i \in [10^{-6},1]$.

\begin{figure}
\resizebox{3.3in}{!}{\includegraphics{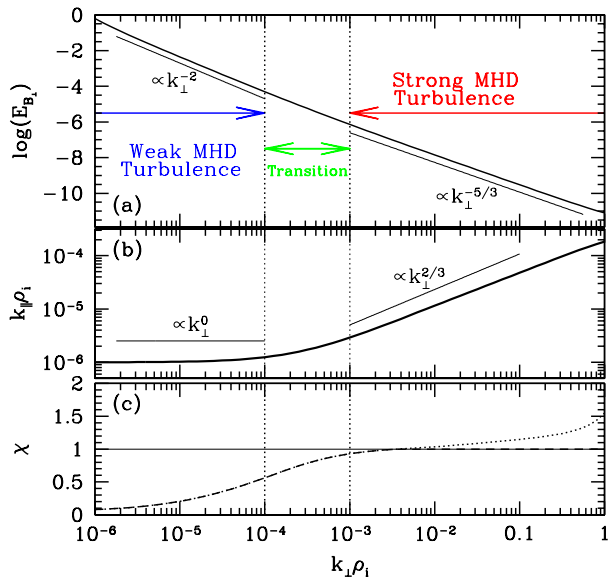}}
\caption{ \label{fig:wsmhd}  Results of  the weakened 
cascade model for the transition from weak MHD turbulence to strong
MHD turbulence for a plasma with $\beta_i=1$ and $T_i/T_e=1$. Plotted
in the panels are: (a) the logarithm of the energy spectrum of the
perpendicular magnetic field fluctuations, $\log(E_{B_\perp}) =\log(
b_k^2/k_\perp)$, normalized by the value of the spectrum
at the driving scale, (b) the normalized parallel wavenumber $k_{\parallel}
\rho_i$, and (c) the nonlinearity parameter $\chi$ 
vs.~the normalized perpendicular wavenumber $k_\perp\rho_i$. Results
show a weak MHD turbulence regime for $k_\perp\rho_i \in
[10^{-6},10^{-4}]$, a transition regime for $k_\perp\rho_i \in
[10^{-4},10^{-3}]$, and a strong MHD turbulence regime for
$k_\perp\rho_i \in [10^{-3},1]$. }
\end{figure}

The results of this test, presented in \figref{fig:wsmhd}, are
consistent with theoretical predictions, showing a transition from the
characteristics of weak MHD turbulence at low perpendicular wavenumber
to those of strong MHD turbulence at high perpendicular
wavenumber. Presented in the panels of \figref{fig:wsmhd} are: (a) the
energy spectrum of the perpendicular magnetic field fluctuations,
$E_{B_\perp} = b_k^2/k_\perp$, normalized by the value of the spectrum
at the driving scale, (b) the normalized parallel wavenumber
$k_{\parallel} \rho_i$, and (c) the nonlinearity parameter $\chi$, all
plotted vs.~the normalized perpendicular wavenumber $k_\perp\rho_i$.
In the weak MHD turbulence regime, over the range of scales $10^{-6} <
k_{\perp } \rho_i < 10^{-4}$, the magnetic energy spectrum yields the
theoretically predicted scaling $E_{B_\perp} \propto k_\perp^{-2}$,
the parallel wavenumber remains essentially constant, and the value of
the nonlinearity parameter is small with  $\chi <0.5$. In the strong MHD
turbulence regime $10^{-3} < k_{\perp } \rho_i < 1$, the magnetic
energy spectrum scales as $E_{B_\perp} \propto k_\perp^{-5/3}$, the
parallel wavenumber scales as predicted by the critical balance
hypothesis $k_\parallel \propto k_\perp^{2/3}$, and the nonlinearity
parameter reaches and maintains a value $\chi \sim 1$. Note that
the nonlinearity parameter is numerically constrained to be $\chi \le
1$ (dashed line); calculating $\chi$ without this constraint yields
the dotted line shown in panel (c), which still remains of order
unity, with $\chi \lesssim 1.5$. Each of these results for the weak
and strong MHD turbulence regimes agree with theoretical
predictions. The range of scales $10^{-4} < k_{\perp } \rho_i <
10^{-3}$ is a transition region that smoothly connects the weak and
strong MHD turbulent cascades.

Given that the state of the art in numerical simulations of turbulence
can reach only a factor of approximately $10^3$ in each dimension,
these results would suggest that it is not currently possible to
capture the physics of the transition from the weak to the strong MHD
turbulence regime in a single numerical simulation. Since the
transition regime spreads over a factor of 10 in dynamic range, and
one would need a minimum dynamic range of 10 in each of the weak and
the strong turbulence regimes, these ranges alone would require all of
the resolution currently feasible, leaving no room for energy
injection and dissipation.  Recent work simulating sub-ranges of this
plot, however, have indeed confirmed the transition from the weak to
the strong scaling of the magnetic energy spectrum in a series of
reduced MHD simulations \cite{Perez:2008}.

\subsection{Local vs.~Nonlocal Models}
To test the importance of the nonlocal contribution to the turbulent
energy cascade rate, we can define a comparable local model by
defining the nonlinear frequency at a given perpendicular wavenumber,
\begin{equation}
 \omega_{nl}(k_\perp) =  C_3 \omega_{nl}^{(\mbox{loc})}( k_\perp).
\label{eq:localwnl}
\end{equation}
Note that the constant $C_3$ does not represent an additional free
parameter for this local model---this constant may be absorbed into
the Kolmogorov constant $C_1$---but is introduced to enable easy
comparison with the nonlocal model given by \eqref{eq:wnl}. The
integration of \eqref{eq:wnl} in the strong MHD and strong KAW
inertial ranges (when damping is negligible), as presented in
Appendix~\ref{app:nonloc_scaling}, suggest that, to yield comparable
local and nonlocal models, the value of this constant should be set to
$C_3=2.25$; in this case, the values of $C_1$ and $C_2$ in the local
and nonlocal models should be directly comparable.  We shall see below
that, although the local model can fit nonlinear numerical simulation
results in certain circumstances, it is missing the essential physics
required to fit a wide range of cases.

In evaluating the nonlocal model, it is desirable to determine the
relative contributions of large-scale shearing, local scale motions,
and small-scale diffusion to the nonlinear frequency at particular
wavenumber, $\omega_{nl}(k_\perp)$. If the definition of the local
contribution given in \eqref{eq:wnl_loc} is used, the delta function
ensures that the local contribution is an infinitesimal slice of the
entire integral, and the local contribution is not easily comparable
to the large- or small-scale contributions. To yield more easily
interpretable results for the contributions to $\omega_{nl}(k_\perp)$,
we split the integral in \eqref{eq:wnl} into three ranges: the
large-scale shearing contribution (s) over $[k_{\perp 0},k_\perp/2)$, the
local-scale contribution (l) over $[k_\perp/2,2k_\perp]$, and the
small-scale diffusive contribution (d)  over $(2k_\perp,k_{\perp max}]$.

Although the weakened cascade model follows only the cascade of
perpendicular magnetic energy, the energy spectra of other fields can
be constructed from the stready-state solution.  Assuming the waves
have the character of the linear Alfv\'enic eigenmodes, we use the
solution of the linear kinetic eigenfunction as a function of
$k_\perp$ to construct, for example, the amplitude of the
perpendicular electric field fluctuation from the amplitude of the
perpendicular magnetic field fluctuation given by the cascade model
solution. Because the phase and amplitude relations between the fields
are fixed by the linear kinetic physics, no additional free parameters
are introduced: if the linear character of the fluctuations applies,
the solution of the perpendicular magnetic energy spectrum determines
the energy spectra of the other fields. This linearity assumption
appears to be well satisfied in comparisons to nonlinear kinetic
simulation results \cite{Howes:2008a}.  This approach
enables us to fit three different curves by adjusting only the two
Kolmogorov constants, $C_1$ and $C_2$, in the weakened cascade model,
providing increased confidence in fits to numerical spectra.

\subsubsection{Moderately Damped $\beta_i=1$ Case}
\begin{figure}
\resizebox{3.0in}{!}{\includegraphics*[0.25in,2.in][7.9in,7.8in]{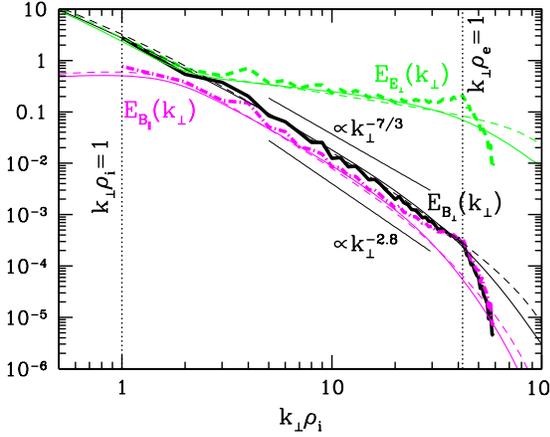}}
\caption{ \label{fig:milestone}  Energy spectra from the $\beta_i=1$
dissipation range simulation \cite{Howes:2011a} resolving both the ion
and electron Larmor radius scales, depicted by vertical dotted
lines. For a $\beta_i=1$ and $T_i/T_e=1$ plasma, thick lines present
numerical energy spectra for the perpendicular magnetic (black solid),
electric (green dashed), and parallel magnetic (magenta dot-dashed)
fields. Predicted energy spectra from the nonlocal model (thin solid)
for $C_1=1.65$ and the local model (thin dashed) for $C_1=1.8$ are
overplotted for comparison.  }
\end{figure}

The first comparison of the local and nonlocal weakened cascade models
tests their ability to fit the results of the $\beta_i=1$ dissipation
range simulation, the first kinetic simulation resolving both the ion
and electron Larmor radius scales in a single simulation, the full details
of which are reported in a companion paper \cite{Howes:2011a}. The
plasma parameters for this gyrokinetic simulation using \T{AstroGK}
\cite{Numata:2010}  are $\beta_i=1$ and $T_i/T_e=1$, and all
dissipation is provided by physically resolved collisionless damping
via the Landau resonance onto the ions and electrons.  In
\figref{fig:milestone} are plotted the one-dimensional energy spectra 
(thick lines) for the perpendicular magnetic field fluctuations
$E_{B_\perp}$ (black solid), the parallel magnetic field fluctuations
$E_{B_\parallel}$ (magenta dot-dashed), and the perpendicular electric
field fluctuations $E_{E_\perp}$ (green dashed). 

For comparison, the predicted energy spectra from the the nonlocal
weakened cascade model (thin solid) and the local cascade model (thin
dashed) are overplotted on \figref{fig:milestone}. Using the linear
gyrokinetic eigenfunctions for the \Alfven mode enables the
determination of the parallel magnetic and perpendicular electric field
spectra from the perpendicular magnetic spectrum output by the cascade
model.  For the nonlocal model, the Kolmogorov constants required to
yield a good fit to the numerical simulation results are $C_1=1.65 \pm
0.20$ and $C_2=1.0$, while for the local model they are $C_1=1.8 \pm
0.35$, $C_2=1.0$, and $C_3=2.25$. Note that a higher value of $C_1$
leads to stronger weighting of the linear damping relative to the
nonlinear energy transfer. The Kolmogorov constant $C_1$ is the
primary adjustable parameter in the weakened cascade model, dominantly
controlling the shape of the energy spectrum.  The second Kolmogorov
constant $C_2$ fine tunes the condition of critical balance, and has
not been adjusted; tests of the distribution of energy in wavevector
space are necessary to constrain this Kolmogorov constant and are
beyond the scope of the present work. 

\begin{figure}
\resizebox{3.0in}{!}{ \includegraphics{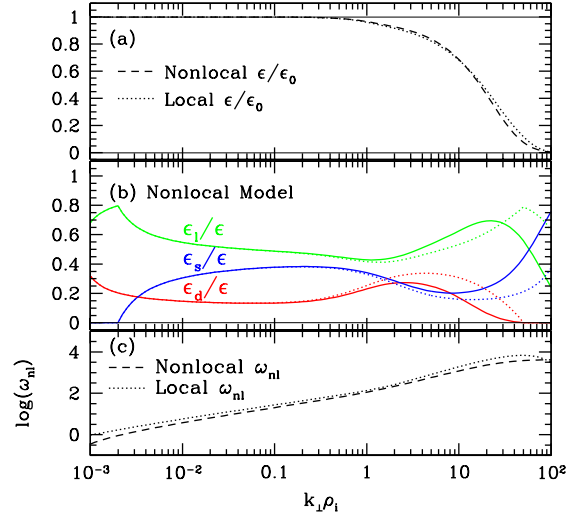}}
\caption{From the local and nonlocal weakened cascade models for the 
 $\beta_i=1$ and $T_i/T_e=1$ plasma depicted in
 \figref{fig:milestone}, (a) the energy cascade rate
 $\epsilon/\epsilon_0$ vs. $k_\perp\rho_i$ for local (dotted) and
 nonlocal (dashed) models, (b) the fractional contribution to
 $\epsilon$ due to the large-scale shearing motions
 $\epsilon_s/\epsilon$ (blue), the local-scale motions
 $\epsilon_l/\epsilon$ (green), and the small-scale diffusive motions
 $\epsilon_d/\epsilon$ (red) with linear kinetic damping (solid) and
 with no damping (dotted), and (c) the nonlinear frequency
 $\omega_{nl}$ from the local model (dotted) and the nonlocal model
 (dashed).
\label{fig:eps_l_sld} }
\end{figure}

Comparison of the energy spectra indicates that both the nonlocal and
local cascade models are able to reproduce the $\beta_i=1$ simulation
spectra with similar values for $C_1$. Differences between the models
become clear as we look more closely at various contributions to the
energy cascade rate, as presented in \figref{fig:eps_l_sld}. In panel
(a), the energy cascade rate $\epsilon$ is plotted vs. $k_\perp\rho_i$
for local (dotted) and nonlocal (dashed) models. This comparison shows
little difference between models, so we must look more closely at the
local and nonlocal contributions to the energy cascade rate.

In panel (b) of \figref{fig:eps_l_sld}, for the nonlocal model, the
fractions of the energy cascade rate from the large-scale shearing
motions $\epsilon_s/\epsilon$ (blue), the local-scale motions
$\epsilon_l/\epsilon$ (green), and the small-scale diffusive motions
$\epsilon_d/\epsilon$ (red) are plotted.  To highlight the effects of
kinetic dissipation on these contributions to the energy cascade rate,
dotted lines give the results when kinetic dissipation is artificially
set to zero.  From this plot, it is clear that, as the cascade
proceeds to higher wavenumber, the diffusive contribution (red) to the
cascade diminishes first at $k_\perp \rho_i
\sim 1$ due to dissipation (compared to the undamped case given by the
red dotted line), leading to a fractional increase in the local and
shear contributions. Next, the fraction of $\epsilon$ due to local
motions (green) begins to diminish at around $k_\perp
\rho_i \sim 20$, leading eventually to a dominance of the energy
cascade rate by the shearing motions of the large scales (blue). It is
this dominance of the energy cascade rate by the large scale motions
as kinetic damping dissipates the turbulent motions that is the
primary difference between the local and nonlocal models. 
 Note that the unusual peak at the
left for $\epsilon_l/\epsilon$ is due to the window that defines the
local contributions \cite{Note3}.

The difference between the local and nonlocal models can be 
seen at high wavenumbers in the nonlinear frequency $\omega_{nl}$,
plotted in panel (c). For the local model (dotted), the nonlinear
frequency peaks at around $k_\perp \rho_i \sim 50$, and then begins to
diminish due to the dissipation of the local motions responsible for
the nonlinear energy transfer. The nonlocal model (dashed), on the
other hand, merely flattens out, as large scale motions continue to
support the nonlinear energy transfer at smaller scales.

In summary, both the local and nonlocal models yield similar results
in modeling the turbulent energy spectra in the moderately damped
$\beta_i=1$ case, as shown in \figref{fig:milestone}. The differences
become apparent only as the kinetic dissipation becomes sufficiently
strong to diminish the local contribution to the nonlinear energy
transfer, enabling nonlocal, large-scale shearing motions to dominate
the nonlinear frequency, as seen at the high $k_\perp \rho_i$ end of
panel (b) in \figref{fig:eps_l_sld}. It is this difference in the
physical mechanisms that will prove crucial in cases where the kinetic
damping is stronger, requiring the additional physics of the effect of
nonlocal motions on the energy transfer to model correctly the
steady-state energy spectra.

\subsubsection{Strongly Damped $\beta_i=0.01$ Case}
\label{sec:strong}

\begin{figure}
\resizebox{3.0in}{!}{ \includegraphics{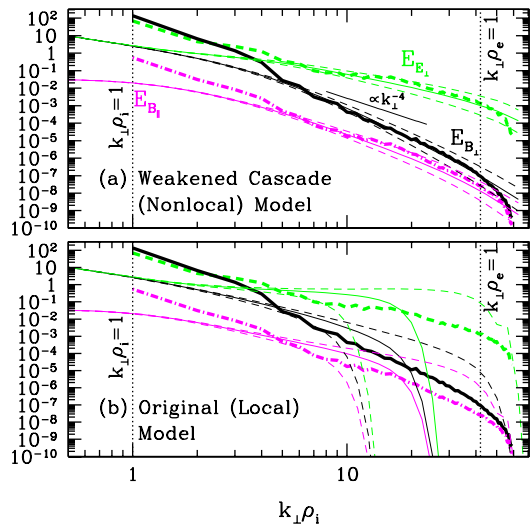}}
\caption{  Energy spectra from the nonlinear \T{AstroGK} gyrokinetic
simulation of turbulence in a plasma with $\beta_i=0.01$ and
$T_i/T_e=1$. Thick lines represent numerical energy spectra for the
perpendicular magnetic (black solid), electric (green dashed), and
parallel magnetic (magenta dot-dashed) fields. (a) The nonlocal model
with $C_1=2.85\pm 0.15$, where thin solid lines are the spectra for
the central value, and the dashed lines demonstrate the $\pm 0.15$
variation.  (b) The local model with $C_1=2.6\pm 0.4$. 
\label{fig:sw42sld} }
\end{figure}

In a low beta plasma, the kinetic damping of fluctuations in the KAW
regime is substantially stronger. In this more strongly damped case,
the difference between the local and nonlocal models is dramatic: the
local model is simply unable to fit the shape of the spectrum. In this
section we compare the spectra predicted by the local and nonlocal
cascade models to the steady state of the $\beta_i=0.01$ nonlinear
\T{AstroGK} simulation. In \figref{fig:sw42sld}, panel (a) shows a fit
of the nonlocal model to the \T{AstroGK} simulation spectra (same
legend as \figref{fig:milestone}), using $C_1=2.85$ (thin solid) and
with $C_1=2.85\pm 0.15$ (thin dashed). In panel (b) is presented the
best fit using the local cascade model, for $C_1=2.6$ (thin solid) and
$C_1=2.6\pm 0.4$ (thin dashed).  All cascade models in this figure use
$C_2=1$ and, as usual, the local model employs $C_3=2.25$ so that the
Kolmogorov constants of both models are comparable. It is immediately
apparent that the local model is incapable of fitting the correct
shape of the spectrum, generally showing power law slopes that are too
flat at the low wavenumbers and a cutoff that is too sharp at high
wavenumbers; the neglect of the effect of nonlocal motions on the
energy cascade rate at high wavenumbers, where the kinetic dissipation
becomes strong, leads to this failure of the local model.

\begin{figure}
\resizebox{3.0in}{!}{\includegraphics*[0.25in,2.in][8.0in,5.9in]{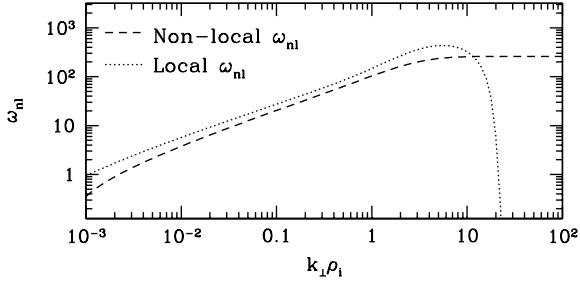}}
\caption{ \label{fig:locvsnonloc}  The  nonlinear frequency $\omega_{nl}$
for the local (dotted) and nonlocal (dashed) cascade models for the
$\beta_i=0.01$ and $T_i/T_e=1$ turbulent plasma shown in \figref{fig:sw42sld}.
}
\end{figure}

An inspection of the nonlinear frequency $\omega_{nl}$ for both
models, plotted in \figref{fig:locvsnonloc}, further illustrates this
point.  The nonlinear frequency for the local model (dotted) peaks at
about $k_\perp \rho_i \simeq 6$, and then drops off rapidly. This
occurs because strong kinetic damping dissipates the turbulent
fluctuations at the local scale, consequently slowing the nonlinear
energy transfer due to those local motions and enabling the linear
kinetic damping to dominate over energy transfer at that scale,
resulting in a sharp cutoff of the turbulent energy spectra. The
nonlocal model (dashed), on the other hand, shows that the nonlinear
frequency flattens to a constant value at high wavenumbers but does
not decrease. In this case, it is the nonlocal, large-scale shearing
motions \cite{Note4}.
that dominate the nonlinear energy transfer
rate at high wavenumbers, leading to turbulent spectra that do not cut
off sharply and are able to fit the nonlinear numerical results. This
evidence suggests that it is the effect of nonlocal motions on the
nonlinear energy transfer rate that is responsible for the more slowly
dissipating, nearly power-law appearance of the turbulent spectra in
both recent solar wind observations
\cite{Sahraoui:2009,Kiyani:2009,Alexandrova:2009,Chen:2010,Sahraoui:2010b} and nonlinear
numerical simulations \cite{Howes:2011a}.

In summary, although the local model can fit the data in certain
moderately damped cases, the effect of nonlocal, large-scale shearing
motions on the nonlinear energy transfer deep in the dissipation range
is essential to avoid a sharp cutoff of the spectra and fit the nearly
power-law behavior observed in both solar wind observations and
nonlinear numerical simulations

\subsubsection{Weak Dissipating KAW Turbulence (WDKT)}
\begin{figure}
\resizebox{3.0in}{!}{ \includegraphics{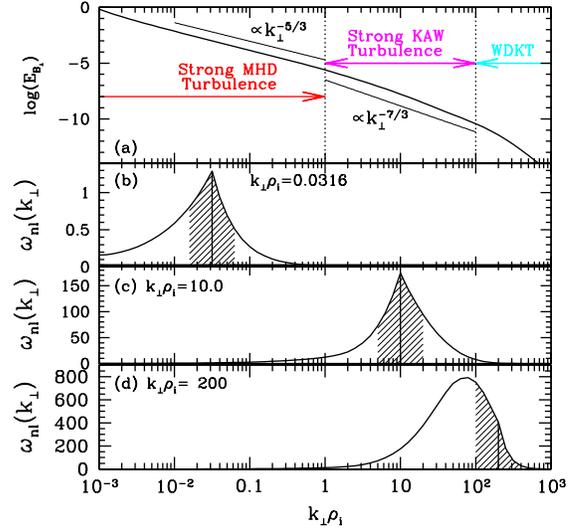}}
\caption{(a) The perpendicular magnetic energy spectrum for strong turbulence
in a plasma with  $\beta_i=1$ and $T_i/T_e=16$ predicted by the weakened
cascade model. In the three lower panels  is plotted the
function in the integrand of \eqref{eq:wnl} for the total nonlinear
frequency $\omega_{nl}(k_\perp)$ at (b) $k_\perp\rho_i = 0.0316$, (c)
$k_\perp\rho_i = 10$, and (d) $k_\perp\rho_i = 200$. The value
of the local wavenumber $k_\perp$ is indicated in each plot by the
vertical solid line, and shading represents contributions due to motions
on local scales from $k_\perp/2$ to $2k_\perp$.
\label{fig:wdkt_wnl} }
\end{figure}

A graphical depiction of the contribution of nonlocal turbulent
fluctuations to the energy cascade rate at a particular wavenumber
$k_\perp$ illuminates the effect of nonlocality. Here the weakened
cascade model, defined in \secref{sec:wc_summary}, is used to solve for
the steady state of a plasma with  $\beta_i=1$ and
$T_i/T_e=16$, parameters chosen so that there is a sufficient dynamic range of
the kinetic
\Alfven wave regime to realize the asymptotic limit of the undamped 
KAW cascade before reaching $k_\perp \rho_e \sim 1$ at $k_\perp \rho_i
\sim 170$.  The model covers a range of wavenumbers $k_\perp\rho_i
\in [10^{-3},10^3]$, employs constants $C_1=4.5$ and $C_2=1$, and is driven
in critical balance with $\chi(k_{\perp 0})=1$.  In
\figref{fig:wdkt_wnl}, the perpendicular magnetic energy spectrum
$E_{B_\perp}$ in panel (a) reveals a strong MHD turbulence regime over
$k_\perp\rho_i \in [10^{-3},1]$ and a strong KAW turbulence regime
over $k_\perp\rho_i \in [1,10^2]$. The kinetic dissipation begins to
significantly alter the spectra for $k_\perp\rho_i \gtrsim 100$,
leading to a transition from strong to weak turbulence locally. But
this situation is not a standard weak turbulence picture, such as
those typically considered in incompressible MHD turbulence. Rather,
the presence of large-scale shearing motions and the importance of
further kinetic dissipation leads to a very different kind of weak KAW
turbulence: we refer to this as the range of \emph{weak dissipating
KAW turbulence} (WDKT), seen over wavenumbers $k_\perp\rho_i \in
[10^2, 10^3]$ in panel (a) of \figref{fig:wdkt_wnl}.

Plotted in the lower three panels of \figref{fig:wdkt_wnl} is the
function in the integrand of \eqref{eq:wnl} for the total nonlinear
frequency $\omega_{nl}(k_\perp)$ at (b) $k_\perp\rho_i = 0.0316$, (c)
$k_\perp\rho_i = 10$, and (d) $k_\perp\rho_i = 200$, where the value
of local wavenumber $k_\perp$ is indicated in each plot by the
vertical solid line. The contributions to the integral yielding
$\omega_{nl}(k_\perp)$ due to local motions is given by the shaded
region below the curve (from $k_\perp/2$ to $2k_\perp$); below the
curve at lower wavenumbers is the contribution from large-scale
shearing motions, and at higher wavenumbers, from small-scale
diffusive motions. In panel (b) is the contribution to the nonlinear
frequency at $k_\perp\rho_i = 0.0316$ in the strong MHD turbulence
regime.  The nonlinear frequency in this regime is dominated by local
motions, with large-scale shearing contributing slightly more than
small-scale diffusive motions. In panel (c) is the contribution to the
nonlinear frequency at $k_\perp\rho_i = 10$ in the strong KAW
turbulence regime. Again, $\omega_{nl}(k_\perp)$ is dominated by local
motions, and here small-scale diffusive motions contribute slightly
more than large-scale shearing. In panel (d) is highlighted one of the
key physics points of the weakened cascade model: the nonlinear
frequency at $k_\perp\rho_i = 200$ in the weak dissipating KAW
turbulence regime is dominated by large-scale shearing motions and not
by local scale motions.  Additionally, there is no contribution from
small-scale diffusion because the cascade is terminated and no motions
exist at smaller scales.


\subsection{Complete Spectrum}
The power of the weakened cascade model is best summarized by a single
final example that demonstrates all of the physics incorporated,
including the transition from weak to strong MHD turbulence and the
complementary transition from strong KAW to weak dissipating KAW
turbulence. This case models the entire turbulent cascade over
$k_\perp\rho_i \in [10^{-5},300]$ for a $\beta_i=1$ and $T_i/T_e=9$
plasma with weak energy injection $\chi(k_{\perp 0})=0.1$ at $k_{\perp
0} \rho_i = k_{\parallel 0} \rho_i =10^{-5}$. The Kolmogorov constants
are $C_1=1.4$ and $C_2=1.0$.

\begin{figure*}
\resizebox{7.0in}{!}{ \includegraphics{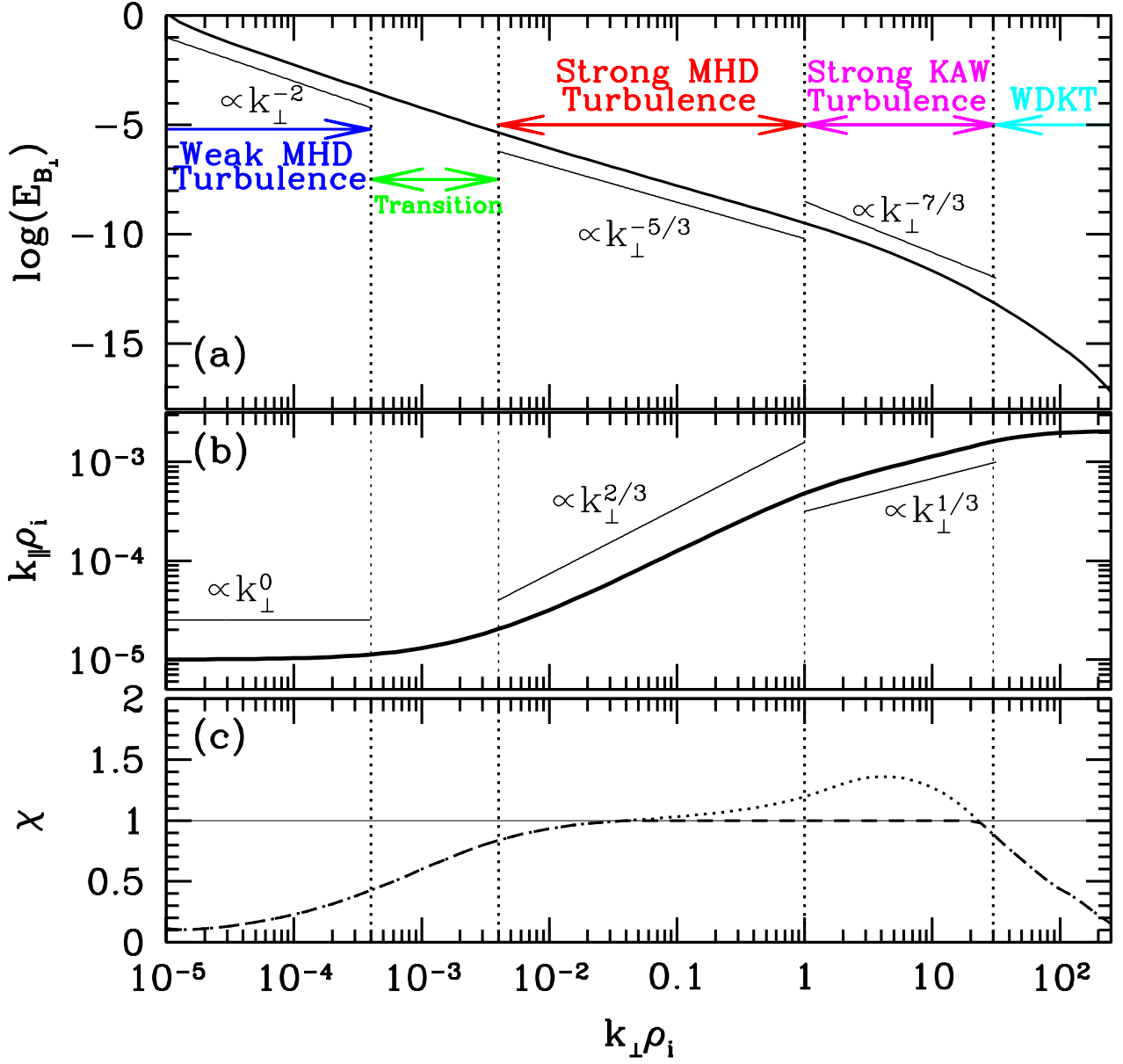}}
\caption{Numerical solution of the weakened cascade model for 
the steady state spectrum in a $\beta_i=1$ and $T_i/T_e=9$ plasma.
The panels plotted are the same as \figref{fig:wsmhd}. Regimes
indicated in panel (a) are weak MHD turbulence, a transition regime
from weak to strong MHD turbulence, strong MHD turbulence, strong KAW
turbulence, and finally weak dissipating KAW turbulence (WDKT).
\label{fig:casc_new} }
\end{figure*}

The steady-state perpendicular magnetic energy spectrum $E_{B_\perp}$
for this case is presented in panel (a) of
\figref{fig:casc_new}. Panel (b) contains the evolution of the
parallel wavenumber $k_\parallel \rho_i$ and panel (c) shows the
evolution of the nonlinearity parameter $\chi$ over the entire
cascade. The solution shows weak MHD turbulence over the range
$k_\perp\rho_i \in [10^{-5},4\times 10^{-4}]$, with a one-dimensional
magnetic energy spectrum $E_{B_\perp} \propto k_\perp ^{-2}$, no
parallel cascade, and weak nonlinearity $\chi < 0.5$. Then comes a regime of transition 
from weak to strong MHD turbulence in the range $k_\perp\rho_i \in [4
\times 10^{-4},4 \times 10^{-3}]$. Over $k_\perp\rho_i \in [4 \times 10^{-3},1]$ is 
strong MHD turbulence with spectrum $E_{B_\perp} \propto k_\perp
^{-5/3}$, parallel cascade scaling as $k_\parallel \propto
k_\perp^{2/3}$ according to critical balance, and a nonlinearity
parameter $\chi \sim 1$. The transition from strong MHD turbulence to
strong KAW turbulence occurs at $k_\perp \rho_i \sim 1$. The range of
strong KAW turbulence over $k_\perp\rho_i \in [1,30]$ yields a
spectrum $E_{B_\perp} \propto k_\perp ^{-7/3}$, parallel cascade
scaling as $k_\parallel \propto k_\perp^{1/3}$ according to critical
balance, and a nonlinearity parameter $\chi \sim 1$.  Finally as the
kinetic dissipation begins to become significant, the cascade becomes
weak dissipating KAW turbulence (WDKT) over $k_\perp\rho_i \in
[30,300]$, with a spectrum that drops off exponentially,
inhibition of the parallel cascade, and a dropping nonlinearity
parameter $\chi < 1$.
\section{Discussion}
The weakened cascade model, summarized in \secref{sec:wc_summary}, is
a refinement of an earlier cascade model \cite{Howes:2008b} intended to 
 explain better the spectra observed in nonlinear numerical simulations
\cite{Howes:2011a} and recent high time resolution observations of the 
dissipation range of turbulence in the near Earth solar wind
\cite{Sahraoui:2009,Kiyani:2009,Alexandrova:2009,Chen:2010,Sahraoui:2010b}. 
The new physical effects incorporated into this model are (1) the
transition between weak and strong turbulence and (2) the effect on
the nonlinear turbulent energy transfer by nonlocal fluctuations.

The model is constructed to have as few free parameters as possible,
letting the linear physics of the kinetic plasma dictate the character
of the turbulent fluctuations. The weakened cascade model has only two
free parameters in the form of dimensionless, order-unity Kolmogorov
constants: $C_1$, which adjusts the weighting of the nonlinear energy
transfer to the linear kinetic damping, and $C_2$, which fine tunes
the condition of critical balance in the nonlinearity parameter
$\chi$. To determine a value of $C_2$ from numerical simulations or
observations is beyond the scope of the present work, so in all cases
we have set $C_2=1$, essentially leaving only a single adjustable
parameter $C_1$ to determine the shape of the steady-state energy
spectra. As presented in \secref{sec:results}, this single degree of
freedom in the weakened cascade model is sufficient to fit closely the
shape of the turbulent spectra from nonlinear gyrokinetic simulations
for both moderately damped $\beta_i=1$ and strongly damped
$\beta_i=0.01$ cases, giving us confidence that the model contains the
essential ingredients necessary to describe successfully the
energetics of the turbulent cascade in a weakly collisional plasma.

Two novel ingredients, inspired by turbulence phenomenology, are
central to the weakened cascade model. The first is the prescription
for the evolution of the parallel cascade depending on the strength of
the nonlinearity parameter, given by \eqref{eq:dkzdkx}. This equation
is constructed to reproduce the predictions of strong and weak
turbulence theories in the appropriate limits.  The second, and likely
more important, ingredient is the nonlocal form of the nonlinear
energy transfer frequency, given by \eqref{eq:wnl}. The abandonment of
the Kolmogorov hypothesis of locality \cite{Kolmogorov:1941} is the
advance necessary to explain the nearly power-law spectra seen in both
the nonlinear simulation results \cite{Howes:2011a} and the solar wind
dissipation range observations
\cite{Sahraoui:2009,Kiyani:2009,Alexandrova:2009,Chen:2010,Sahraoui:2010b}. 
When dissipation begins to weaken
the local scale motions, the smaller scale motions are necessarily yet
more strongly damped, so diffusion by those smaller scale motions is
effectively negligible. Therefore, the key result of the weakened
cascade model presented here is that the nonlocal effect of shearing by large scale
motions explains the nearly power law appearance of numerical and
observational spectra in the dissipation range (this effect is
particularly critical in the low beta case presented in
\secref{sec:strong}).

The locality of MHD turbulence in the inertial range has gained a
significant amount of attention in the literature recently, with some
studies finding evidence for nonlocality
\cite{Schekochihin:2004d,Alexakis:2005,Carati:2006,Yousef:2007}, while 
other studies claim locality in the asymptotic limit of large
Reynold's numbers \cite{Aluie:2010} (see Mininni 2010
\cite{Mininni:2010} for a recent review of this topic). At question 
is whether the properties of the MHD inertial range become independent
of the details of the mechanisms responsible for driving and
dissipating the turbulence, enabling the determination of a
self-similar solution for a universal MHD turbulent spectrum.  When
the dissipation range of plasma turbulence is being considered,
however, it is obvious that no self-similar solution can be found, so
the importance of a nonlocal contribution to the energy cascade rate
in the dissipation range, as proposed here, should come as no
surprise.

It is interesting to note that scale locality would be recovered, and
therefore a self-similar solution would arise, if the kinetic \Alfven
wave range of scales, $\rho_i \ll 1/k \ll \rho_e$, becomes
asymptotically large, as occurs in the limit of vanishing electron
mass, $m_e \rightarrow 0$. In this very small electron mass limit $m_e
\rightarrow 0$, the kinetic \Alfven wave parallel phase velocity is
always much less than the electron thermal velocity,
$\omega/k_\parallel \ll v_{te}$, so collisionless damping via resonant
wave-particle interactions with the electrons becomes negligible. In
the absence of damping, a second inertial range is recovered for the
dispersive kinetic \Alfven waves over the range of scales $\rho_i \ll
1/k \ll \rho_e$. The weakened cascade model describes this limit
correctly, because if $m_e \rightarrow 0$, then the collisionless
damping rate $\gamma\rightarrow 0$ in \eqref{eq:damping}.  In this
case, the weakened cascade model recovers a second inertial range of
kinetic \Alfven waves with a spectral index of $-7/3$, as shown by the
dashed line denoted the ``undamped model'' in Figure~3 of the original
cascade model paper by Howes \emph{et al.}  (2008) \cite{Howes:2008b}.

\subsection{Comparison to Previous Cascade Models}
It is instructive to compare the predictions of the weakened cascade
model, with its inclusion of nonlocal effects on the energy cascade
rate, to those of the original cascade model \cite{Howes:2008b} and of
a similar model by Podesta \emph{et al.} 2010\cite{Podesta:2010a},
both strictly local models.  Two of the general conclusions of the
original cascade model were that variations in collisionless damping
with plasma parameters naturally explained observed variations in
solar wind turbulent spectra and that the dissipation range spectrum
should be an exponential fall off
\cite{Howes:2008b}. Podesta \emph{et al.}~make the stronger claim that
``an energy cascade consisting solely of KAWs cannot reach scales of
order the electron gyroradius, $k_\perp \rho_e \sim 1$. It implies
that the power-law spectrum in the regime of electron scales must be
supported by wave modes other than the KAW'' \cite{Podesta:2010a}.

Since the only wave mode supported in the dissipation range of a
$\beta_i=1$ plasma in gyrokinetic theory is the kinetic \Alfven wave,
the $\beta_i=1$ gyrokinetic simulation \cite{Howes:2011a} demonstrates
unequivocally by counterexample that the claim by Podesta \emph{et
al.}~\cite{Podesta:2010a}---that the kinetic \Alfven wave cascade
cannot reach scales $k_\perp \rho_e \sim 1$---is incorrect. We explain
below that the primary reason for this failure lies in the
determination of the Kolmogorov constant in their cascade model.

Each of these cascade models incorporates an order-unity dimensionless
constant that essentially adjusts the weighting of nonlinear energy
transfer to that of the linear kinetic damping.  Podesta \emph{et al.}~estimate $k_\perp \rho_i \simeq 28$ as the wavenumber at which the KAW
cascade terminates, given by their equation~(23), which is linearly
dependent on their constant $A$. Without any guidance to determine
this constant, Podesta \emph{et al.}~base their conclusions on the
theory with $A=1$. We can usefully compare their model with the local
version of the weakened cascade model (using
equation~[\ref{eq:localwnl}] for the nonlinear frequency) to determine what value of 
$A$ would fit the $\beta_i=1$ dissipation range simulation spectra in
\figref{fig:milestone}. Using $b_k^2= k E(k)$ in the limit $k_\perp \gg k_\parallel$ to connect
to their model, the energy cascade rate $\epsilon$ in our local model
is given by
\begin{equation}
\epsilon = C_1^{-3/2}  C_3 \omega_{nl}^{(\mbox{loc})}  
k E(k).
\label{eq:localeps}
\end{equation}
The energy cascade rate $\varepsilon$ in Podesta \emph{et al.}, based
on their equation~(6) \cite{Note5},
is given by
\begin{equation}
\varepsilon = \frac{A} {2 \pi} \omega k E(k).
\end{equation}
The assumption of critical balance is taken by Podesta \emph{et
al.}~to be $\omega = \omega_{nl}^{(\mbox{loc})}$, so the two models
should give similar results if $A=2 \pi C_1^{-3/2} C_3$.  Substituting
in the best fit values $C_1=1.8$ and $C_3=2.25$ from
\figref{fig:milestone}, we find $A = 5.85$. This comparison suggests
that Podesta \emph{et al.}~significantly underestimated the weight of
the energy cascade rate relative to the linear kinetic damping,
leading to a conclusion inconsistent with direct numerical simulations
of kinetic turbulence.  Nonlinear kinetic simulations of turbulence
therefore play an invaluable role in the effort to understand
turbulence in the solar wind and other weakly collisional
astrophysical plasmas.

\subsection{Limitations of the Weakened Cascade Model}
The weakened cascade model has been developed to predict the turbulent
energy spectra occurring in weakly collisional astrophysical
plasmas. We discuss here a number of assumptions that have been made in the
construction of the model.

First, the model has been constructed to reproduce the scaling of the
turbulent energy spectra given by the Goldreich and Sridhar theories
for weak and strong MHD turbulence
\cite{Sridhar:1994,Goldreich:1995,Lithwick:2003}, and  their extension to
turbulence in the kinetic \Alfven wave regime. An alternative theory
for the scaling of strong MHD turbulence has been proposed by Boldyrev
\cite{Boldyrev:2006}, and substantial numerical support  has accumulated 
in its favor
\cite{Mason:2006,Mason:2008,Perez:2008,Boldyrev:2009,Perez:2009,Perez:2010}.
Modifications of the weakened cascade model to reproduce instead the Bolydrev
scalings will be discussed in a subsequent paper.

Another potential limitation of the weakened cascade model is the conjecture
that the parallel cascade of energy is inhibited for weak turbulence in 
\emph{both} the MHD and KAW regimes, as discussed in \secref{sec:weak}.

The one-dimensional nature of the weakened cascade model restricts its
direct applicability to plasmas in which there is a single scale of
energy injection at the outer scale $(k_{\perp 0}, k_{\parallel
0})$. As discussed in Howes \emph{et al.} 2008 \cite{Howes:2008b}, the
lack of structure at small scales in the solar wind energy spectrum is
evidence against significant energy injection as scales smaller than
the outer scale $(k_{\perp 0}, k_{\parallel 0})$, so the model appears
to be broadly applicable to the solar wind.

A number of other factors that may significantly affect the turbulent
fluctuations in the solar wind are not incorporated into the weakened
cascade model, including the radial expansion of the solar wind,
kinetic temperature anisotropy instabilities, and the imbalance of
sunward and anti-sunward \Alfven wave energy fluxes.  A detailed
discussion of these effects is presented in the paper describing the
original cascade model \cite{Howes:2008b} and will not be repeated
here.  It suffices to say that the weakened cascade model is an
attempt to understand the quantitative details of the energy transport
in balanced, Alfv\'enic turbulence in a weakly collisional plasma; the
relation of these additional effects to this fundamental turbulent 
evolution merits further investigation.

The weakened cascade model does not account for one recently
discovered physical mechanism that may play an important role in the
energy transport in weakly collisional Alfv\'enic turbulence: the
entropy
cascade\cite{Schekochihin:2009,Tatsuno:2009,Plunk:2010,Tatsuno:2010,Plunk:2011}.
As described in Schekochihin \emph{et al.}~2009
\cite{Schekochihin:2009}, the ion entropy cascade is a dual cascade to
small scales in both physical space and velocity space of the ion
distribution function.  Operating at scales below the ion Larmor
radius $k_\perp \rho_i > 1$, in the regime of kinetic \Alfven wave
turbulence, this process is driven by nonlinear phase mixing and
represents an alternative channel of energy transport that is not
included in the weakened cascade model.  Comparisons of the
predictions of the weakened cascade model with the results of a suite
of nonlinear gyrokinetic simulations will enable an evaluation of the
importance of the ion entropy cascade in the turbulent energy transport,
an important line of future research.

\section{Conclusions}
Early cascade models for kinetic turbulence in weakly collisional
astrophysical plasmas, such as the solar wind, suggested that the
energy spectra in the dissipation range should exponentially fall off
\cite{Howes:2008b}, and that due to collisionless damping, kinetic
\Alfven waves could not be responsible for the cascade of
energy to electron scales \cite{Podesta:2010a}. However, nonlinear
kinetic simulations of turbulence over the entire dissipation range
\cite{Howes:2011a} and high time resolution observations of the
dissipation range fluctuations in the solar wind
\cite{Sahraoui:2009,Kiyani:2009,Alexandrova:2009,Chen:2010,Sahraoui:2010b} yield nearly 
power-law energy spectra rather than an exponential decay. The failure
of these early cascade models motivated refinements to explain the
numerical and observational results, yielding the 
\emph{weakened cascade model} presented here.

The original cascade model by Howes \emph{et al.}
2008\cite{Howes:2008b} was based on three assumptions: (1) the
Kolmogorov hypothesis of the locality of the nonlinear energy transfer
in wavenumber space, (2) the conjecture of critical balance at all
scales, and (3) the applicability of linear kinetic damping rates. The
weakened cascade model eliminates the first two assumptions of the
original model, resulting in a more broadly applicable and physically
realistic model.

The assumption of critical balance at all scales is dropped; instead,
the model handles explicitly the transition between weak and strong
turbulence. A key point in the treatment of weak turbulence is our
conjecture that the parallel cascade is inhibited in both the MHD and
kinetic \Alfven wave regimes. This physics is contained in the model
for the evolution of the parallel wavenumber given by
\eqref{eq:dkzdkx}.  The more significant advance of the weakened
cascade model is the abandonment of the locality hypothesis of
Kolmogorov. Although energy is still transferred locally in wavenumber
space---for example, from $k_\perp$ to $2 k_\perp$---the turbulent
fluctuations responsible for that energy transfer may be nonlocal.
Both shearing by motions at larger scales and diffusion by motions at
smaller scales contribute to the nonlinear energy cascade rate
according to \eqref{eq:wnl}. 

In \secref{sec:results}, we have demonstrated that the weakened
cascade model reproduces the transition from weak to strong MHD
turbulence as predicted by theory. As the collisionless dissipation in
the kinetic \Alfven wave regime becomes significant, the model also
shows a complementary transition from strong KAW turbulence to
\emph{weak dissipating KAW turbulence}, a new regime of weak
turbulence in which the effect of shearing by large scale motions and
continued kinetic dissipation play an important role.

The key result of this paper is that the nearly power-law energy
spectra observed in the dissipation range of both numerical
simulations and solar wind observations are explained by the inclusion
of the effect of nonlocal motions on the nonlinear energy cascade
rate, specifically the shearing by large-scale motions. Although
numerical spectra for a moderately damped $\beta_i=1$ plasma may be
equally well explained by either local or nonlocal models
(\figref{fig:milestone}), for the more strongly damped $\beta_i=0.01$
plasma, the inclusion of nonlocal effects is critical for the model to
fit the numerical energy spectra (\figref{fig:sw42sld}). The
importance of the nonlocal shearing motions to the energy cascade rate
is demonstrated in panel (d) of
\figref{fig:wdkt_wnl}, where it is clear that the large-scale
contribution dominates over the local contribution. The effect at
these strongly dissipative scales is that the nonlinear frequency does
not decrease with increasing perpendicular wavenumber, as a local
model would suggest, but that it remains constant due to the
large-scale contribution, as shown in \figref{fig:locvsnonloc}. Thus,
by abandoning the Kolmogorov hypothesis of locality, the weakened
cascade model explains the nearly power-law spectra found in numerical
and observational studies of the dissipation range by including the
nonlocal effect of large-scale shearing motions on the energy transfer
rate.

The ultimate aim of the weakened cascade model is not to fit numerical
and observational turbulent spectra, but to predict them. Nonlinear
kinetic simulations of dissipation range turbulence have already
played an important role in evaluating the weakened cascade
model. Since the Kolmogorov constants $C_1$ and $C_2$ may, in
principle, depend on the plasma parameters $\beta_i$ and $T_i/T_e$,
numerical studies will continue to play a critical role as we perform
a suite of kinetic turbulence simulations over a range of plasma
parameters to determine this dependence, ultimately striving for
predictive capability. In addition, these numerical studies will also
be crucial in judging the importance of the ion entropy cascade to the
turbulent energy transfer in the dissipation range.



%
%

%

\begin{acknowledgments}
G.G.H. thanks Alex Schekochihin and Eliot Quataert for encouragement
and insightful discussions.  This work was supported by the DOE Center
for Multiscale Plasma Dynamics, STFC, Leverhulme Trust Network for
 Magnetised Plasma Turbulence, NSF CAREER Award AGS-1054061, and NASA
NNX10AC91G. Computing resources were supplied through DOE INCITE Award
PSS002, NSF TeraGrid Award PHY090084, and DOE INCITE Award FUS030.
\end{acknowledgments}

\appendix

\section{The Effect of Nonlocal Interactions on Turbulence Scaling}
\label{app:nonloc_scaling}
Here we consider the effect of the nonlocal contribution to the
nonlinear frequency, given by \eqref{eq:wnl}, on the scaling of the
spectra in the strong MHD and strong KAW inertial ranges.

In the strong MHD inertial range, the contribution to the nonlinear
frequency due to motions at each wavenumber is
$\omega_{nl}^{(\mbox{loc})}= k_\perp b_k = C_1^{1/2} \epsilon^{1/3}
k_\perp^{2/3}$. Performing the integral in \eqref{eq:wnl} to find
$\omega_{nl}$, we obtain
\begin{eqnarray}
 \omega_{nl}(k_\perp) & = &  \frac{3}{2}C_1^{1/2} \epsilon^{1/3} \left( k_\perp^{2/3} - k_{\perp 0}^{2/3}\right) \nonumber \\
&+& \frac{3}{4}C_1^{1/2} \epsilon^{1/3}k_\perp^2 \left(k_\perp^{-4/3} - k_{\perp max}^{-4/3}\right),
\end{eqnarray}
where the first term is due to the large-scale shearing, and the
second is due to the small-scale diffusion.  The inertial range is
defined as the range of scales unaffected by large-scale driving or
small-scale dissipation, corresponding to the limit $k_{\perp 0} \ll
k_\perp \ll k_{\perp max}$.  In this limit, the nonlinear
frequency simplifies to 
\begin{equation}
 \omega_{nl}(k_\perp)  =   \frac{3}{2}\omega_{nl}^{(\mbox{loc})} + 
 \frac{3}{4}\omega_{nl}^{(\mbox{loc})} = \frac{9}{4}\omega_{nl}^{(\mbox{loc})} (k_\perp).
\end{equation}
Therefore, within the strong MHD inertial range, the nonlinear model
yields a nonlinear frequency of the form $ \omega_{nl}(k_\perp) = C_3
\omega_{nl}^{(\mbox{loc})}( k_\perp)$, where the constant is $C_3=2.25$.  
This order-unity constant factor is the only difference between
models, so the scaling of the nonlocal model within the strong MHD
inertial range will therefore be the same as a local model. Note that
the contribution of the large-scale shearing motions is twice that of
the small-scale diffusive motions.

In the strong KAW inertial range, we have $\omega_{nl}^{(\mbox{loc})}=
k_\perp b_k k_\perp \rho_i= C_1^{1/2} \epsilon^{1/3} \rho_i^{2/3}
k_\perp^{4/3}$.  Performing the integral in \eqref{eq:wnl}, we find
\begin{eqnarray}
 \omega_{nl}(k_\perp) & = &  \frac{3}{4}C_1^{1/2} \epsilon^{1/3} \rho_i^{2/3} \left( k_\perp^{4/3} - k_{\perp 0}^{4/3}\right) \nonumber \\
&+& \frac{3}{2}C_1^{1/2} \epsilon^{1/3} \rho_i^{2/3}  k_\perp^2 \left(k_\perp^{-2/3} - k_{\perp max}^{-2/3}\right),
\end{eqnarray}
where the first term is the large-scale contribution and the second is
the small-scale contribution.  Within the KAW inertial range, we apply
the limit $k_{\perp 0} \ll k_\perp \ll k_{\perp max}$ to simplify 
the nonlinear frequency to
\begin{equation}
 \omega_{nl}(k_\perp) = \frac{3}{4}\omega_{nl}^{(\mbox{loc})} +
 \frac{3}{2}\omega_{nl}^{(\mbox{loc})} =
 \frac{9}{4}\omega_{nl}^{(\mbox{loc})} (k_\perp).
\end{equation}
Again, we find that the scaling of the nonlocal model in the strong
KAW inertial range will be that same as that of a local models, the
only difference being the same constant factor $C_3=2.25$.  In this case,
however, the small-scale diffusive motions contribute twice as much to
the nonlinear frequency as the large-scale shearing motions.


\begin{thebibliography}{10}%
\makeatletter
\providecommand \@ifxundefined [1]{%
 \ifx #1\undefined \expandafter \@firstoftwo
 \else \expandafter \@secondoftwo
\fi
}%
\providecommand \@ifnum [1]{%
 \ifnum #1\expandafter \@firstoftwo
 \else \expandafter \@secondoftwo
\fi
}%
\providecommand \enquote [1]{``#1''}%
\providecommand \bibnamefont  [1]{#1}%
\providecommand \bibfnamefont [1]{#1}%
\providecommand \citenamefont [1]{#1}%
\providecommand\href[0]{\@sanitize\@href}%
\providecommand\@href[1]{\endgroup\@@startlink{#1}\endgroup\@@href}%
\providecommand\@@href[1]{#1\@@endlink}%
\providecommand \@sanitize [0]{\begingroup\catcode`\&12\catcode`\#12\relax}%
\@ifxundefined \pdfoutput {\@firstoftwo}{%
 \@ifnum{\z@=\pdfoutput}{\@firstoftwo}{\@secondoftwo}%
}{%
 \providecommand\@@startlink[1]{\leavevmode}%
 \providecommand\@@endlink[0]{}%
}{%
 \providecommand\@@startlink[1]{%
  \leavevmode
  \pdfstartlink
   attr{/Border[0 0 1 ]/H/I/C[0 1 1]}%
   user{/Subtype/Link/A<</Type/Action/S/URI/URI(#1)>>}%
  \relax
 }%
 \providecommand\@@endlink[0]{\pdfendlink}%
}%
\providecommand \url  [0]{\begingroup\@sanitize \@url }%
\providecommand \@url [1]{\endgroup\@href {#1}{\urlprefix}}%
\providecommand \urlprefix [0]{URL }%
\providecommand \Eprint[0]{\href }%
\@ifxundefined \urlstyle {%
  \providecommand \doi [1]{doi:\discretionary{}{}{}#1}%
}{%
  \providecommand \doi [0]{doi:\discretionary{}{}{}\begingroup
  \urlstyle{rm}\Url }%
}%
\providecommand \doibase [0]{http://dx.doi.org/}%
\providecommand \Doi[1]{\href{\doibase#1}}%
\providecommand \selectlanguage [0]{\@gobble}%
\providecommand \bibinfo [0]{\@secondoftwo}%
\providecommand \bibfield [0]{\@secondoftwo}%
\providecommand \translation [1]{[#1]}%
\providecommand \BibitemOpen[0]{}%
\providecommand \bibitemStop [0]{}%
\providecommand \bibitemNoStop [0]{.\EOS\space}%
\providecommand \EOS [0]{\spacefactor3000\relax}%
\providecommand \BibitemShut [1]{\csname bibitem#1\endcsname}%
\bibitem{Coleman:1968}%
  \BibitemOpen
  \bibfield{author}{%
  \bibinfo {author} {\bibfnamefont{P.~J.}\ \bibnamefont{{Coleman}},
  \bibfnamefont{Jr.}},\ }%
  \bibfield{title}{%
  \enquote{\bibinfo {title} {{Turbulence, Viscosity, and Dissipation in the
  Solar-Wind Plasma}},}\ }%
  \bibfield{journal}{%
  \bibinfo {journal} {Astrophys.~J.}\ }%
  \textbf{\bibinfo {volume} {153}},\ \bibinfo {pages} {371--388} (\bibinfo
  {month} {Aug.}\ \bibinfo {year} {1968})\BibitemShut{NoStop}%
\bibitem{Marsch:1991}%
  \BibitemOpen
  \bibfield{author}{%
  \bibinfo {author} {\bibfnamefont{E.}~\bibnamefont{{Marsch}}},\ }%
  \enquote{\bibinfo {title} {{Kinetic Physics of the Solar Wind Plasma}},}\ in\
  \emph{\bibinfo {booktitle} {{Physics of the Inner Heliosphere II. Particles,
  Waves and Turbulence.}}},\ \bibinfo {editor} {edited by\ \bibinfo {editor}
  {\bibfnamefont{E}~\bibnamefont{{Schwenn}}, \bibfnamefont{R.~and{Marsch}}}}\
  (\bibinfo {publisher} {{Springer-Verlag}},\ \bibinfo {address} {{Berlin}},\
  \bibinfo {year} {1991})\ pp.\ \bibinfo {pages} {45--133}\BibitemShut{NoStop}%
\bibitem{Marsch:2006}%
  \BibitemOpen
  \bibfield{author}{%
  \bibinfo {author} {\bibfnamefont{E.}~\bibnamefont{Marsch}},\ }%
  \bibfield{title}{%
  \enquote{\bibinfo {title} {Kinetic physics of the solar corona and solar
  wind},}\ }%
  \bibfield{journal}{%
  \bibinfo {journal} {Living Rev.~Solar Phys.}\ }%
  \textbf{\bibinfo {volume} {3}} (\bibinfo {year} {2006}),\
  \url{http://www.livingreviews.org/lrsp-2006-1}\BibitemShut{NoStop}%
\bibitem{Howes:2008c}%
  \BibitemOpen
  \bibfield{author}{%
  \bibinfo {author} {\bibfnamefont{G.~G.}\ \bibnamefont{{Howes}}},\ }%
  \bibfield{title}{%
  \enquote{\bibinfo {title} {{Inertial range turbulence in kinetic plasmas}},}\
  }%
  \bibfield{journal}{%
  \Doi{10.1063/1.2889005}{\bibinfo {journal} {Phys.~Plasmas}}\ }%
  \textbf{\bibinfo {volume} {15}},\ \bibinfo {pages} {055904} (\bibinfo {month}
  {May}\ \bibinfo {year} {2008})\BibitemShut{NoStop}%
\bibitem{Howes:2008b}%
  \BibitemOpen
  \bibfield{author}{%
  \bibinfo {author} {\bibfnamefont{G.~G.}\ \bibnamefont{{Howes}}}, \bibinfo
  {author} {\bibfnamefont{S.~C.}\ \bibnamefont{{Cowley}}}, \bibinfo {author}
  {\bibfnamefont{W.}~\bibnamefont{{Dorland}}}, \bibinfo {author}
  {\bibfnamefont{G.~W.}\ \bibnamefont{{Hammett}}}, \bibinfo {author}
  {\bibfnamefont{E.}~\bibnamefont{{Quataert}}},\ and\ \bibinfo {author}
  {\bibfnamefont{A.~A.}\ \bibnamefont{{Schekochihin}}},\ }%
  \bibfield{title}{%
  \enquote{\bibinfo {title} {{A model of turbulence in magnetized plasmas:
  Implications for the dissipation range in the solar wind}},}\ }%
  \bibfield{journal}{%
  \Doi{10.1029/2007JA012665}{\bibinfo {journal} {J.~Geophys.~Res.}}\ }%
  \textbf{\bibinfo {volume} {113}},\ \bibinfo {pages} {A05103} (\bibinfo
  {month} {May}\ \bibinfo {year} {2008}),\
  \Eprint{http://arxiv.org/abs/arXiv:0707.3147}{arXiv:0707.3147}\BibitemShut{N%
oStop}%
\bibitem{Howes:2008d}%
  \BibitemOpen
  \bibfield{author}{%
  \bibinfo {author} {\bibfnamefont{G.~G.}\ \bibnamefont{{Howes}}}, \bibinfo
  {author} {\bibfnamefont{S.~C.}\ \bibnamefont{{Cowley}}}, \bibinfo {author}
  {\bibfnamefont{W.}~\bibnamefont{{Dorland}}}, \bibinfo {author}
  {\bibfnamefont{G.~W.}\ \bibnamefont{{Hammett}}}, \bibinfo {author}
  {\bibfnamefont{E.}~\bibnamefont{{Quataert}}}, \bibinfo {author}
  {\bibfnamefont{A.~A.}\ \bibnamefont{{Schekochihin}}},\ and\ \bibinfo {author}
  {\bibfnamefont{T.}~\bibnamefont{{Tatsuno}}},\ }%
  \bibfield{title}{%
  \enquote{\bibinfo {title} {{Howes et al. Reply:}},}\ }%
  \bibfield{journal}{%
  \Doi{10.1103/PhysRevLett.101.149502}{\bibinfo {journal} {Phys.~Rev.~Lett.}}\
  }%
  \textbf{\bibinfo {volume} {101}},\ \bibinfo {pages} {149502} (\bibinfo
  {month} {Oct.}\ \bibinfo {year} {2008})\BibitemShut{NoStop}%
\bibitem{Schekochihin:2009}%
  \BibitemOpen
  \bibfield{author}{%
  \bibinfo {author} {\bibfnamefont{A.~A.}\ \bibnamefont{{Schekochihin}}},
  \bibinfo {author} {\bibfnamefont{S.~C.}\ \bibnamefont{{Cowley}}}, \bibinfo
  {author} {\bibfnamefont{W.}~\bibnamefont{{Dorland}}}, \bibinfo {author}
  {\bibfnamefont{G.~W.}\ \bibnamefont{{Hammett}}}, \bibinfo {author}
  {\bibfnamefont{G.~G.}\ \bibnamefont{{Howes}}}, \bibinfo {author}
  {\bibfnamefont{E.}~\bibnamefont{{Quataert}}},\ and\ \bibinfo {author}
  {\bibfnamefont{T.}~\bibnamefont{{Tatsuno}}},\ }%
  \bibfield{title}{%
  \enquote{\bibinfo {title} {{Astrophysical Gyrokinetics: Kinetic and Fluid
  Turbulent Cascades in Magnetized Weakly Collisional Plasmas}},}\ }%
  \bibfield{journal}{%
  \Doi{10.1088/0067-0049/182/1/310}{\bibinfo {journal} {Astrophys.~J.~Supp.}}\
  }%
  \textbf{\bibinfo {volume} {182}},\ \bibinfo {pages} {310--377} (\bibinfo
  {month} {May}\ \bibinfo {year} {2009})\BibitemShut{NoStop}%
\bibitem{Howes:2008a}%
  \BibitemOpen
  \bibfield{author}{%
  \bibinfo {author} {\bibfnamefont{G.~G.}\ \bibnamefont{{Howes}}}, \bibinfo
  {author} {\bibfnamefont{W.}~\bibnamefont{{Dorland}}}, \bibinfo {author}
  {\bibfnamefont{S.~C.}\ \bibnamefont{{Cowley}}}, \bibinfo {author}
  {\bibfnamefont{G.~W.}\ \bibnamefont{{Hammett}}}, \bibinfo {author}
  {\bibfnamefont{E.}~\bibnamefont{{Quataert}}}, \bibinfo {author}
  {\bibfnamefont{A.~A.}\ \bibnamefont{{Schekochihin}}},\ and\ \bibinfo {author}
  {\bibfnamefont{T.}~\bibnamefont{{Tatsuno}}},\ }%
  \bibfield{title}{%
  \enquote{\bibinfo {title} {{Kinetic Simulations of Magnetized Turbulence in
  Astrophysical Plasmas}},}\ }%
  \bibfield{journal}{%
  \Doi{10.1103/PhysRevLett.100.065004}{\bibinfo {journal} {Phys.~Rev.~Lett.}}\
  }%
  \textbf{\bibinfo {volume} {100}},\ \bibinfo {pages} {065004} (\bibinfo
  {month} {Feb.}\ \bibinfo {year} {2008})\BibitemShut{NoStop}%
\bibitem{Howes:2011a}%
  \BibitemOpen
  \bibfield{author}{%
  \bibinfo {author} {\bibfnamefont{G.~G.}\ \bibnamefont{Howes}}, \bibinfo
  {author} {\bibfnamefont{J.~M.}\ \bibnamefont{TenBarge}}, \bibinfo {author}
  {\bibfnamefont{W.}~\bibnamefont{Dorland}}, \bibinfo {author}
  {\bibfnamefont{E.}~\bibnamefont{Quataert}}, \bibinfo {author}
  {\bibfnamefont{A.~A.}\ \bibnamefont{Schekochihin}}, \bibinfo {author}
  {\bibfnamefont{R.}~\bibnamefont{Numata}},\ and\ \bibinfo {author}
  {\bibfnamefont{T.}~\bibnamefont{Tatsuno}},\ }%
  \bibfield{title}{%
  \enquote{\bibinfo {title} {Gyrokinetic simulations of solar wind turbulence
  from ion to electron scales},}\ }%
  \bibfield{journal}{%
  \Doi{10.1103/PhysRevLett.107.035004}{\bibinfo {journal} {Phys.~Rev.~Lett.}}\
  }%
  \textbf{\bibinfo {volume} {107}},\ \bibinfo {pages} {035004} (\bibinfo
  {month} {Jul}\ \bibinfo {year} {2011})\BibitemShut{NoStop}%
\bibitem{Sahraoui:2009}%
  \BibitemOpen
  \bibfield{author}{%
  \bibinfo {author} {\bibfnamefont{F.}~\bibnamefont{{Sahraoui}}}, \bibinfo
  {author} {\bibfnamefont{M.~L.}\ \bibnamefont{{Goldstein}}}, \bibinfo {author}
  {\bibfnamefont{P.}~\bibnamefont{{Robert}}},\ and\ \bibinfo {author}
  {\bibfnamefont{Y.~V.}\ \bibnamefont{{Khotyaintsev}}},\ }%
  \bibfield{title}{%
  \enquote{\bibinfo {title} {{Evidence of a Cascade and Dissipation of
  Solar-Wind Turbulence at the Electron Gyroscale}},}\ }%
  \bibfield{journal}{%
  \Doi{10.1103/PhysRevLett.102.231102}{\bibinfo {journal} {Phys.~Rev.~Lett.}}\
  }%
  \textbf{\bibinfo {volume} {102}},\ \bibinfo {pages} {231102--+} (\bibinfo
  {month} {Jun.}\ \bibinfo {year} {2009})\BibitemShut{NoStop}%
\bibitem{Kiyani:2009}%
  \BibitemOpen
  \bibfield{author}{%
  \bibinfo {author} {\bibfnamefont{K.~H.}\ \bibnamefont{{Kiyani}}}, \bibinfo
  {author} {\bibfnamefont{S.~C.}\ \bibnamefont{{Chapman}}}, \bibinfo {author}
  {\bibfnamefont{Y.~V.}\ \bibnamefont{{Khotyaintsev}}}, \bibinfo {author}
  {\bibfnamefont{M.~W.}\ \bibnamefont{{Dunlop}}},\ and\ \bibinfo {author}
  {\bibfnamefont{F.}~\bibnamefont{{Sahraoui}}},\ }%
  \bibfield{title}{%
  \enquote{\bibinfo {title} {{Global scale-invariant dissipation in
  collisionless plasma turbulence}},}\ }%
  \bibfield{journal}{%
  \bibinfo {journal} {Phys.~Rev.~Lett.}\ }%
  \textbf{\bibinfo {volume} {103}},\ \bibinfo {pages} {075006} (\bibinfo {year}
  {2009})\BibitemShut{NoStop}%
\bibitem{Alexandrova:2009}%
  \BibitemOpen
  \bibfield{author}{%
  \bibinfo {author} {\bibfnamefont{O.}~\bibnamefont{{Alexandrova}}}, \bibinfo
  {author} {\bibfnamefont{J.}~\bibnamefont{{Saur}}}, \bibinfo {author}
  {\bibfnamefont{C.}~\bibnamefont{{Lacombe}}}, \bibinfo {author}
  {\bibfnamefont{A.}~\bibnamefont{{Mangeney}}}, \bibinfo {author}
  {\bibfnamefont{J.}~\bibnamefont{{Mitchell}}}, \bibinfo {author}
  {\bibfnamefont{S.~J.}\ \bibnamefont{{Schwartz}}},\ and\ \bibinfo {author}
  {\bibfnamefont{P.}~\bibnamefont{{Robert}}},\ }%
  \bibfield{title}{%
  \enquote{\bibinfo {title} {{Universality of Solar-Wind Turbulent Spectrum
  from MHD to Electron Scales}},}\ }%
  \bibfield{journal}{%
  \Doi{10.1103/PhysRevLett.103.165003}{\bibinfo {journal} {Phys.~Rev.~Lett.}}\
  }%
  \textbf{\bibinfo {volume} {103}},\ \bibinfo {pages} {165003--+} (\bibinfo
  {month} {Oct.}\ \bibinfo {year} {2009})\BibitemShut{NoStop}%
\bibitem{Chen:2010}%
  \BibitemOpen
  \bibfield{author}{%
  \bibinfo {author} {\bibfnamefont{C.~H.~K.}\ \bibnamefont{{Chen}}}, \bibinfo
  {author} {\bibfnamefont{T.~S.}\ \bibnamefont{{Horbury}}}, \bibinfo {author}
  {\bibfnamefont{A.~A.}\ \bibnamefont{{Schekochihin}}}, \bibinfo {author}
  {\bibfnamefont{R.~T.}\ \bibnamefont{{Wicks}}}, \bibinfo {author}
  {\bibfnamefont{O.}~\bibnamefont{{Alexandrova}}},\ and\ \bibinfo {author}
  {\bibfnamefont{J.}~\bibnamefont{{Mitchell}}},\ }%
  \bibfield{title}{%
  \enquote{\bibinfo {title} {{Anisotropy of Solar Wind Turbulence between Ion
  and Electron Scales}},}\ }%
  \bibfield{journal}{%
  \Doi{10.1103/PhysRevLett.104.255002}{\bibinfo {journal} {Physical Review
  Letters}}\ }%
  \textbf{\bibinfo {volume} {104}},\ \bibinfo {pages} {255002--+} (\bibinfo
  {month} {Jun.}\ \bibinfo {year} {2010}),\
  \Eprint{http://arxiv.org/abs/1002.2539}{arXiv:1002.2539
  [physics.space-ph]}\BibitemShut{NoStop}%
\bibitem{Sahraoui:2010b}%
  \BibitemOpen
  \bibfield{author}{%
  \bibinfo {author} {\bibfnamefont{F.}~\bibnamefont{{Sahraoui}}}, \bibinfo
  {author} {\bibfnamefont{M.~L.}\ \bibnamefont{{Goldstein}}}, \bibinfo {author}
  {\bibfnamefont{G.}~\bibnamefont{{Belmont}}}, \bibinfo {author}
  {\bibfnamefont{P.}~\bibnamefont{{Canu}}},\ and\ \bibinfo {author}
  {\bibfnamefont{L.}~\bibnamefont{{Rezeau}}},\ }%
  \bibfield{title}{%
  \enquote{\bibinfo {title} {{Three Dimensional Anisotropic k Spectra of
  Turbulence at Subproton Scales in the Solar Wind}},}\ }%
  \bibfield{journal}{%
  \Doi{10.1103/PhysRevLett.105.131101}{\bibinfo {journal} {Phys.~Rev.~Lett.}}\
  }%
  \textbf{\bibinfo {volume} {105}},\ \bibinfo {pages} {131101--+} (\bibinfo
  {month} {Sep.}\ \bibinfo {year} {2010})\BibitemShut{NoStop}%
\bibitem{Note0}%
  \BibitemOpen
  \bibinfo {note} {At spacecraft-frame frequencies above about $0.4$~Hz,
  \emph{in situ} measurements of the solar wind one-dimensional magnetic energy
  frequency spectrum show that the spectrum becomes steeper than the $-5/3$
  spectral index typically found at lower frequencies \cite{Bale:2005}. The
  range of frequencies at and above this break in the spectrum has
  traditionally been denoted the \emph{dissipation range} of solar wind
  turbulence. In the present work, the dissipation range is interpreted to be
  associated with the perpendicular scale of the ion Larmor radius, $k_\perp
  \rho_i \gtrsim 1$. Other studies have argued for different terminology to
  describe this range, such as dispersion range \cite{Stawicki:2001} or
  scattering range \cite{Rudakov:2011}. Here, we choose to employ the more
  traditional terminology and denote the range of scales $k_\perp \rho_i
  \gtrsim 1$ as the dissipation range.}\BibitemShut{Stop}%
\bibitem{Kolmogorov:1941}%
  \BibitemOpen
  \bibfield{author}{%
  \bibinfo {author} {\bibfnamefont{A.~N.}\ \bibnamefont{Kolmogorov}},\ }%
  \bibfield{title}{%
  \enquote{\bibinfo {title} {The local structure of turbulence in
  incompressible viscous fluid for very large reynolds numbers},}\ }%
  \bibfield{journal}{%
  \bibinfo {journal} {Dokl. Akad. Nauk SSSR}\ }%
  \textbf{\bibinfo {volume} {30}},\ \bibinfo {pages} {9--} (\bibinfo {year}
  {1941}),\ \bibinfo {note} {english Translation: Proc. Roy. Soc. London A,
  434, 9 (1991)}\BibitemShut{NoStop}%
\bibitem{Iroshnikov:1963}%
  \BibitemOpen
  \bibfield{author}{%
  \bibinfo {author} {\bibfnamefont{R.~S.}\ \bibnamefont{Iroshnikov}},\ }%
  \bibfield{title}{%
  \enquote{\bibinfo {title} {The turbulence of a conducting fluid in a strong
  magnetic field},}\ }%
  \bibfield{journal}{%
  \bibinfo {journal} {Astron. Zh.}\ }%
  \textbf{\bibinfo {volume} {40}},\ \bibinfo {pages} {742} (\bibinfo {year}
  {1963}),\ \bibinfo {note} {{English} Translation: Sov. Astron., 7 566
  (1964)}\BibitemShut{NoStop}%
\bibitem{Kraichnan:1965}%
  \BibitemOpen
  \bibfield{author}{%
  \bibinfo {author} {\bibfnamefont{R.~H.}\ \bibnamefont{Kraichnan}},\ }%
  \bibfield{title}{%
  \enquote{\bibinfo {title} {Inertial range spectrum of hyromagnetic
  turbulence},}\ }%
  \bibfield{journal}{%
  \bibinfo {journal} {Phys.~Fluids}\ }%
  \textbf{\bibinfo {volume} {8}},\ \bibinfo {pages} {1385--1387} (\bibinfo
  {year} {1965})\BibitemShut{NoStop}%
\bibitem{Robinson:1971}%
  \BibitemOpen
  \bibfield{author}{%
  \bibinfo {author} {\bibfnamefont{D.~C.}\ \bibnamefont{Robinson}}\ and\
  \bibinfo {author} {\bibfnamefont{M.~G.}\ \bibnamefont{Rusbridge}},\ }%
  \bibfield{title}{%
  \enquote{\bibinfo {title} {Structure of turbulence in the zeta plasma},}\ }%
  \bibfield{journal}{%
  \bibinfo {journal} {Phys.~Fluids}\ }%
  \textbf{\bibinfo {volume} {14}},\ \bibinfo {pages} {2499--2511} (\bibinfo
  {year} {1971})\BibitemShut{NoStop}%
\bibitem{Zweben:1979}%
  \BibitemOpen
  \bibfield{author}{%
  \bibinfo {author} {\bibfnamefont{S.~J.}\ \bibnamefont{Zweben}}, \bibinfo
  {author} {\bibfnamefont{C.~R.}\ \bibnamefont{Menyuk}},\ and\ \bibinfo
  {author} {\bibfnamefont{R.~J.}\ \bibnamefont{Taylor}},\ }%
  \bibfield{title}{%
  \enquote{\bibinfo {title} {Small-scale magnetic fluctuations inside the
  macrotor tokamak},}\ }%
  \bibfield{journal}{%
  \bibinfo {journal} {Phys.~Rev.~Lett.}\ }%
  \textbf{\bibinfo {volume} {42}},\ \bibinfo {pages} {1270--1274} (\bibinfo
  {year} {1979})\BibitemShut{NoStop}%
\bibitem{Montgomery:1981}%
  \BibitemOpen
  \bibfield{author}{%
  \bibinfo {author} {\bibfnamefont{D.}~\bibnamefont{Montgomery}}\ and\ \bibinfo
  {author} {\bibfnamefont{L.}~\bibnamefont{Turner}},\ }%
  \bibfield{title}{%
  \enquote{\bibinfo {title} {Anisotropic magnetohydrodynamic turbulence in a
  strong external magnetic field},}\ }%
  \bibfield{journal}{%
  \bibinfo {journal} {Phys.~Fluids}\ }%
  \textbf{\bibinfo {volume} {24}},\ \bibinfo {pages} {825--831} (\bibinfo
  {year} {1981})\BibitemShut{NoStop}%
\bibitem{Belcher:1971}%
  \BibitemOpen
  \bibfield{author}{%
  \bibinfo {author} {\bibfnamefont{J.~W.}\ \bibnamefont{{Belcher}}}\ and\
  \bibinfo {author} {\bibfnamefont{L.}~\bibnamefont{{Davis}}},\ }%
  \bibfield{title}{%
  \enquote{\bibinfo {title} {{Large-Amplitude Alfv{\'e}n Waves in the
  Interplanetary Medium, 2}},}\ }%
  \bibfield{journal}{%
  \bibinfo {journal} {J.~Geophys.~Res.}\ }%
  \textbf{\bibinfo {volume} {76}},\ \bibinfo {pages} {3534--3563} (\bibinfo
  {year} {1971})\BibitemShut{NoStop}%
\bibitem{Shebalin:1983}%
  \BibitemOpen
  \bibfield{author}{%
  \bibinfo {author} {\bibfnamefont{J.~V.}\ \bibnamefont{Shebalin}}, \bibinfo
  {author} {\bibfnamefont{W.~H.}\ \bibnamefont{Matthaeus}},\ and\ \bibinfo
  {author} {\bibfnamefont{D.}~\bibnamefont{Montgomery}},\ }%
  \bibfield{title}{%
  \enquote{\bibinfo {title} {Anisotropy in mhd turbulence due to a mean
  magnetic field},}\ }%
  \bibfield{journal}{%
  \bibinfo {journal} {J.~Plasma Phys.}\ }%
  \textbf{\bibinfo {volume} {29}},\ \bibinfo {pages} {525--547} (\bibinfo
  {year} {1983})\BibitemShut{NoStop}%
\bibitem{Sridhar:1994}%
  \BibitemOpen
  \bibfield{author}{%
  \bibinfo {author} {\bibfnamefont{S.}~\bibnamefont{Sridhar}}\ and\ \bibinfo
  {author} {\bibfnamefont{P.}~\bibnamefont{Goldreich}},\ }%
  \bibfield{title}{%
  \enquote{\bibinfo {title} {{Toward a Thoery of Interstellar Turbulence I.
  Weak Alfv\'enic Turbulence}},}\ }%
  \bibfield{journal}{%
  \bibinfo {journal} {Astrophys.~J.}\ }%
  \textbf{\bibinfo {volume} {433}},\ \bibinfo {pages} {612--621} (\bibinfo
  {year} {1994})\BibitemShut{NoStop}%
\bibitem{Goldreich:1995}%
  \BibitemOpen
  \bibfield{author}{%
  \bibinfo {author} {\bibfnamefont{P.}~\bibnamefont{Goldreich}}\ and\ \bibinfo
  {author} {\bibfnamefont{S.}~\bibnamefont{Sridhar}},\ }%
  \bibfield{title}{%
  \enquote{\bibinfo {title} {{Toward a Thoery of Interstellar Turbulence II.
  Strong Alfv\'enic Turbulence}},}\ }%
  \bibfield{journal}{%
  \bibinfo {journal} {Astrophys.~J.}\ }%
  \textbf{\bibinfo {volume} {438}},\ \bibinfo {pages} {763--775} (\bibinfo
  {year} {1995})\BibitemShut{NoStop}%
\bibitem{Montgomery:1995}%
  \BibitemOpen
  \bibfield{author}{%
  \bibinfo {author} {\bibfnamefont{D.}~\bibnamefont{{Montgomery}}}\ and\
  \bibinfo {author} {\bibfnamefont{W.~H.}\ \bibnamefont{{Matthaeus}}},\ }%
  \bibfield{title}{%
  \enquote{\bibinfo {title} {{Anisotropic Modal Energy Transfer in Interstellar
  Turbulence}},}\ }%
  \bibfield{journal}{%
  \Doi{10.1086/175910}{\bibinfo {journal} {Astrophys.~J.}}\ }%
  \textbf{\bibinfo {volume} {447}},\ \bibinfo {pages} {706--+} (\bibinfo
  {month} {Jul.}\ \bibinfo {year} {1995})\BibitemShut{NoStop}%
\bibitem{Ng:1996}%
  \BibitemOpen
  \bibfield{author}{%
  \bibinfo {author} {\bibfnamefont{C.~S.}\ \bibnamefont{{Ng}}}\ and\ \bibinfo
  {author} {\bibfnamefont{A.}~\bibnamefont{{Bhattacharjee}}},\ }%
  \bibfield{title}{%
  \enquote{\bibinfo {title} {{Interaction of Shear-Alfven Wave Packets:
  Implication for Weak Magnetohydrodynamic Turbulence in Astrophysical
  Plasmas}},}\ }%
  \bibfield{journal}{%
  \Doi{10.1086/177468}{\bibinfo {journal} {Astrophys.~J.}}\ }%
  \textbf{\bibinfo {volume} {465}},\ \bibinfo {pages} {845--+} (\bibinfo
  {month} {Jul.}\ \bibinfo {year} {1996})\BibitemShut{NoStop}%
\bibitem{Goldreich:1997}%
  \BibitemOpen
  \bibfield{author}{%
  \bibinfo {author} {\bibfnamefont{P.}~\bibnamefont{Goldreich}}\ and\ \bibinfo
  {author} {\bibfnamefont{S.}~\bibnamefont{Sridhar}},\ }%
  \bibfield{title}{%
  \enquote{\bibinfo {title} {Magnetohydrodynamic turbulence revisited},}\ }%
  \bibfield{journal}{%
  \bibinfo {journal} {Astrophys.~J.}\ }%
  \textbf{\bibinfo {volume} {485}},\ \bibinfo {pages} {680--688} (\bibinfo
  {year} {1997})\BibitemShut{NoStop}%
\bibitem{Ng:1997}%
  \BibitemOpen
  \bibfield{author}{%
  \bibinfo {author} {\bibfnamefont{C.~S.}\ \bibnamefont{{Ng}}}\ and\ \bibinfo
  {author} {\bibfnamefont{A.}~\bibnamefont{{Bhattacharjee}}},\ }%
  \bibfield{title}{%
  \enquote{\bibinfo {title} {{Scaling of anisotropic spectra due to the weak
  interaction of shear-Alfv{\'e}n wave packets}},}\ }%
  \bibfield{journal}{%
  \bibinfo {journal} {Phys.~Plasmas}\ }%
  \textbf{\bibinfo {volume} {4}},\ \bibinfo {pages} {605--610} (\bibinfo
  {month} {Mar.}\ \bibinfo {year} {1997})\BibitemShut{NoStop}%
\bibitem{Galtier:2000}%
  \BibitemOpen
  \bibfield{author}{%
  \bibinfo {author} {\bibfnamefont{S.}~\bibnamefont{{Galtier}}}, \bibinfo
  {author} {\bibfnamefont{S.~V.}\ \bibnamefont{{Nazarenko}}}, \bibinfo {author}
  {\bibfnamefont{A.~C.}\ \bibnamefont{{Newell}}},\ and\ \bibinfo {author}
  {\bibfnamefont{A.}~\bibnamefont{{Pouquet}}},\ }%
  \bibfield{title}{%
  \enquote{\bibinfo {title} {{A weak turbulence theory for incompressible
  magnetohydrodynamics}},}\ }%
  \bibfield{journal}{%
  \bibinfo {journal} {J.~Plasma Phys.}\ }%
  \textbf{\bibinfo {volume} {63}},\ \bibinfo {pages} {447--488} (\bibinfo
  {month} {Jun.}\ \bibinfo {year} {2000}),\
  \Eprint{http://arxiv.org/abs/astro-ph/0008148}{astro-ph/0008148}\BibitemShut%
{NoStop}%
\bibitem{Lithwick:2003}%
  \BibitemOpen
  \bibfield{author}{%
  \bibinfo {author} {\bibfnamefont{Y.}~\bibnamefont{Lithwick}}\ and\ \bibinfo
  {author} {\bibfnamefont{P.}~\bibnamefont{Goldreich}},\ }%
  \bibfield{title}{%
  \enquote{\bibinfo {title} {Imbalanced weak magnetohydrodynamic turbulence},}\
  }%
  \bibfield{journal}{%
  \bibinfo {journal} {Astrophys.~J.}\ }%
  \textbf{\bibinfo {volume} {582}},\ \bibinfo {pages} {1220--1240} (\bibinfo
  {year} {2003})\BibitemShut{NoStop}%
\bibitem{Perez:2008}%
  \BibitemOpen
  \bibfield{author}{%
  \bibinfo {author} {\bibfnamefont{J.~C.}\ \bibnamefont{{Perez}}}\ and\
  \bibinfo {author} {\bibfnamefont{S.}~\bibnamefont{{Boldyrev}}},\ }%
  \bibfield{title}{%
  \enquote{\bibinfo {title} {{On Weak and Strong Magnetohydrodynamic
  Turbulence}},}\ }%
  \bibfield{journal}{%
  \Doi{10.1086/526342}{\bibinfo {journal} {Astrophys.~J.~Lett.}}\ }%
  \textbf{\bibinfo {volume} {672}},\ \bibinfo {pages} {L61--L64} (\bibinfo
  {month} {Jan.}\ \bibinfo {year} {2008}),\
  \Eprint{http://arxiv.org/abs/0712.2086}{arXiv:0712.2086}\BibitemShut{NoStop}%
\bibitem{Higdon:1984a}%
  \BibitemOpen
  \bibfield{author}{%
  \bibinfo {author} {\bibfnamefont{J.~C.}\ \bibnamefont{Higdon}},\ }%
  \bibfield{title}{%
  \enquote{\bibinfo {title} {Density fluctuations in the interstellar medium:
  Evidence for anisotropic magnetogasdynamic turbulence i. model and
  astrophysical sites},}\ }%
  \bibfield{journal}{%
  \bibinfo {journal} {Astrophys.~J.}\ }%
  \textbf{\bibinfo {volume} {285}},\ \bibinfo {pages} {109--123} (\bibinfo
  {year} {1984})\BibitemShut{NoStop}%
\bibitem{Cho:2000}%
  \BibitemOpen
  \bibfield{author}{%
  \bibinfo {author} {\bibfnamefont{J.}~\bibnamefont{Cho}}\ and\ \bibinfo
  {author} {\bibfnamefont{E.~T.}\ \bibnamefont{Vishniac}},\ }%
  \bibfield{title}{%
  \enquote{\bibinfo {title} {{The Anisotropy of Magnetohydrodynamic Alfv\'enic
  Turbulence}},}\ }%
  \bibfield{journal}{%
  \bibinfo {journal} {Astrophys.~J.}\ }%
  \textbf{\bibinfo {volume} {539}},\ \bibinfo {pages} {273--282} (\bibinfo
  {year} {2000})\BibitemShut{NoStop}%
\bibitem{Maron:2001}%
  \BibitemOpen
  \bibfield{author}{%
  \bibinfo {author} {\bibfnamefont{J.}~\bibnamefont{Maron}}\ and\ \bibinfo
  {author} {\bibfnamefont{P.}~\bibnamefont{Goldreich}},\ }%
  \bibfield{title}{%
  \enquote{\bibinfo {title} {Simulations of incompressible magnetohydrodynamic
  turbulence},}\ }%
  \bibfield{journal}{%
  \bibinfo {journal} {Astrophys.~J.}\ }%
  \textbf{\bibinfo {volume} {554}},\ \bibinfo {pages} {1175--1196} (\bibinfo
  {year} {2001})\BibitemShut{NoStop}%
\bibitem{Horbury:2008}%
  \BibitemOpen
  \bibfield{author}{%
  \bibinfo {author} {\bibfnamefont{T.~S.}\ \bibnamefont{{Horbury}}}, \bibinfo
  {author} {\bibfnamefont{M.}~\bibnamefont{Forman}},\ and\ \bibinfo {author}
  {\bibfnamefont{S.}~\bibnamefont{Oughton}},\ }%
  \bibfield{title}{%
  \enquote{\bibinfo {title} {Anisotropic scaling of magnetohydrodynamic
  turbulence},}\ }%
  \bibfield{journal}{%
  \Doi{10.1103/PhysRevLett.101.175005}{\bibinfo {journal} {Phys.~Rev.~Lett.}}\
  }%
  \textbf{\bibinfo {volume} {101}},\ \bibinfo {pages} {175005} (\bibinfo
  {month} {Oct}\ \bibinfo {year} {2008})\BibitemShut{NoStop}%
\bibitem{Podesta:2009a}%
  \BibitemOpen
  \bibfield{author}{%
  \bibinfo {author} {\bibfnamefont{J.~J.}\ \bibnamefont{{Podesta}}},\ }%
  \bibfield{title}{%
  \enquote{\bibinfo {title} {{Dependence of Solar-Wind Power Spectra on the
  Direction of the Local Mean Magnetic Field}},}\ }%
  \bibfield{journal}{%
  \Doi{10.1088/0004-637X/698/2/986}{\bibinfo {journal} {Astrophys.~J.}}\ }%
  \textbf{\bibinfo {volume} {698}},\ \bibinfo {pages} {986--999} (\bibinfo
  {month} {Jun.}\ \bibinfo {year} {2009}),\
  \Eprint{http://arxiv.org/abs/0901.4940}{arXiv:0901.4940}\BibitemShut{NoStop}%
\bibitem{Gruzinov:1998}%
  \BibitemOpen
  \bibfield{author}{%
  \bibinfo {author} {\bibfnamefont{A.~V.}\ \bibnamefont{{Gruzinov}}},\ }%
  \bibfield{title}{%
  \enquote{\bibinfo {title} {{Radiative Efficiency of Collisionless
  Accretion}},}\ }%
  \bibfield{journal}{%
  \Doi{10.1086/305845}{\bibinfo {journal} {Astrophys.~J.}}\ }%
  \textbf{\bibinfo {volume} {501}},\ \bibinfo {pages} {787--+} (\bibinfo
  {month} {Jul.}\ \bibinfo {year} {1998}),\
  \Eprint{http://arxiv.org/abs/astro-ph/9710132}{astro-ph/9710132}\BibitemShut%
{NoStop}%
\bibitem{Quataert:1999}%
  \BibitemOpen
  \bibfield{author}{%
  \bibinfo {author} {\bibfnamefont{E.}~\bibnamefont{{Quataert}}}\ and\ \bibinfo
  {author} {\bibfnamefont{A.}~\bibnamefont{{Gruzinov}}},\ }%
  \bibfield{title}{%
  \enquote{\bibinfo {title} {{Turbulence and Particle Heating in
  Advection-dominated Accretion Flows}},}\ }%
  \bibfield{journal}{%
  \Doi{10.1086/307423}{\bibinfo {journal} {Astrophys.~J.}}\ }%
  \textbf{\bibinfo {volume} {520}},\ \bibinfo {pages} {248--255} (\bibinfo
  {month} {Jul.}\ \bibinfo {year} {1999}),\
  \Eprint{http://arxiv.org/abs/astro-ph/9803112}{astro-ph/9803112}\BibitemShut%
{NoStop}%
\bibitem{Biskamp:1999}%
  \BibitemOpen
  \bibfield{author}{%
  \bibinfo {author} {\bibfnamefont{D.}~\bibnamefont{Biskmap}}, \bibinfo
  {author} {\bibfnamefont{E.}~\bibnamefont{Schwarz}}, \bibinfo {author}
  {\bibfnamefont{A.}~\bibnamefont{Zeiler}}, \bibinfo {author}
  {\bibfnamefont{A.}~\bibnamefont{Celani}},\ and\ \bibinfo {author}
  {\bibfnamefont{J.~F.}\ \bibnamefont{Drake}},\ }%
  \bibfield{title}{%
  \enquote{\bibinfo {title} {Electron magnetohydrodynamic turbulence},}\ }%
  \bibfield{journal}{%
  \bibinfo {journal} {Phys.~Plasmas}\ }%
  \textbf{\bibinfo {volume} {6}},\ \bibinfo {pages} {751--758} (\bibinfo {year}
  {1999})\BibitemShut{NoStop}%
\bibitem{Cho:2004}%
  \BibitemOpen
  \bibfield{author}{%
  \bibinfo {author} {\bibfnamefont{J.}~\bibnamefont{{Cho}}}\ and\ \bibinfo
  {author} {\bibfnamefont{A.}~\bibnamefont{{Lazarian}}},\ }%
  \bibfield{title}{%
  \enquote{\bibinfo {title} {{The Anisotropy of Electron Magnetohydrodynamic
  Turbulence}},}\ }%
  \bibfield{journal}{%
  \Doi{10.1086/425215}{\bibinfo {journal} {Astrophys.~J.~Lett.}}\ }%
  \textbf{\bibinfo {volume} {615}},\ \bibinfo {pages} {L41--L44} (\bibinfo
  {month} {Nov.}\ \bibinfo {year} {2004}),\
  \Eprint{http://arxiv.org/abs/astro-ph/0406595}{astro-ph/0406595}\BibitemShut%
{NoStop}%
\bibitem{Krishan:2004}%
  \BibitemOpen
  \bibfield{author}{%
  \bibinfo {author} {\bibfnamefont{V.}~\bibnamefont{{Krishan}}}\ and\ \bibinfo
  {author} {\bibfnamefont{S.~M.}\ \bibnamefont{{Mahajan}}},\ }%
  \bibfield{title}{%
  \enquote{\bibinfo {title} {{Magnetic fluctuations and Hall
  magnetohydrodynamic turbulence in the solar wind}},}\ }%
  \bibfield{journal}{%
  \Doi{10.1029/2004JA010496}{\bibinfo {journal} {J.~Geophys.~Res.}}\ }%
  \textbf{\bibinfo {volume} {109}},\ \bibinfo {pages} {A11105} (\bibinfo
  {month} {Nov.}\ \bibinfo {year} {2004})\BibitemShut{NoStop}%
\bibitem{Dastgeer:2005}%
  \BibitemOpen
  \bibfield{author}{%
  \bibinfo {author} {\bibfnamefont{D.}~\bibnamefont{{Shaikh}}}\ and\ \bibinfo
  {author} {\bibfnamefont{G.~P.}\ \bibnamefont{{Zank}}},\ }%
  \bibfield{title}{%
  \enquote{\bibinfo {title} {{Driven dissipative whistler wave turbulence}},}\
  }%
  \bibfield{journal}{%
  \Doi{10.1063/1.2146957}{\bibinfo {journal} {Phys.~Plasmas}}\ }%
  \textbf{\bibinfo {volume} {12}},\ \bibinfo {pages} {2310--+} (\bibinfo
  {month} {Dec.}\ \bibinfo {year} {2005})\BibitemShut{NoStop}%
\bibitem{Cho:2002}%
  \BibitemOpen
  \bibfield{author}{%
  \bibinfo {author} {\bibfnamefont{J.}~\bibnamefont{{Cho}}}, \bibinfo {author}
  {\bibfnamefont{A.}~\bibnamefont{{Lazarian}}},\ and\ \bibinfo {author}
  {\bibfnamefont{E.~T.}\ \bibnamefont{{Vishniac}}},\ }%
  \bibfield{title}{%
  \enquote{\bibinfo {title} {{Simulations of Magnetohydrodynamic Turbulence in
  a Strongly Magnetized Medium}},}\ }%
  \bibfield{journal}{%
  \Doi{10.1086/324186}{\bibinfo {journal} {Astrophys.~J.}}\ }%
  \textbf{\bibinfo {volume} {564}},\ \bibinfo {pages} {291--301} (\bibinfo
  {month} {Jan.}\ \bibinfo {year} {2002}),\
  \Eprint{http://arxiv.org/abs/astro-ph/0105235}{astro-ph/0105235}\BibitemShut%
{NoStop}%
\bibitem{Cho:2003}%
  \BibitemOpen
  \bibfield{author}{%
  \bibinfo {author} {\bibfnamefont{J.}~\bibnamefont{{Cho}}}\ and\ \bibinfo
  {author} {\bibfnamefont{A.}~\bibnamefont{{Lazarian}}},\ }%
  \bibfield{title}{%
  \enquote{\bibinfo {title} {{Compressible magnetohydrodynamic turbulence: mode
  coupling, scaling relations, anisotropy, viscosity-damped regime and
  astrophysical implications}},}\ }%
  \bibfield{journal}{%
  \Doi{10.1046/j.1365-8711.2003.06941.x}{\bibinfo {journal}
  {Mon.~Not.~Roy.~Astron.~Soc.}}\ }%
  \textbf{\bibinfo {volume} {345}},\ \bibinfo {pages} {325--339} (\bibinfo
  {month} {Oct.}\ \bibinfo {year} {2003}),\
  \Eprint{http://arxiv.org/abs/astro-ph/0301062}{astro-ph/0301062}\BibitemShut%
{NoStop}%
\bibitem{Oughton:2004}%
  \BibitemOpen
  \bibfield{author}{%
  \bibinfo {author} {\bibfnamefont{S.}~\bibnamefont{{Oughton}}}, \bibinfo
  {author} {\bibfnamefont{P.}~\bibnamefont{{Dmitruk}}},\ and\ \bibinfo {author}
  {\bibfnamefont{W.~H.}\ \bibnamefont{{Matthaeus}}},\ }%
  \bibfield{title}{%
  \enquote{\bibinfo {title} {{Reduced magnetohydrodynamics and parallel
  spectral transfer}},}\ }%
  \bibfield{journal}{%
  \Doi{10.1063/1.1705652}{\bibinfo {journal} {Phys.~Plasmas}}\ }%
  \textbf{\bibinfo {volume} {11}},\ \bibinfo {pages} {2214--2225} (\bibinfo
  {month} {May}\ \bibinfo {year} {2004})\BibitemShut{NoStop}%
\bibitem{Boldyrev:2006}%
  \BibitemOpen
  \bibfield{author}{%
  \bibinfo {author} {\bibfnamefont{S.}~\bibnamefont{{Boldyrev}}},\ }%
  \bibfield{title}{%
  \enquote{\bibinfo {title} {{Spectrum of Magnetohydrodynamic Turbulence}},}\
  }%
  \bibfield{journal}{%
  \Doi{10.1103/PhysRevLett.96.115002}{\bibinfo {journal} {Phys.~Rev.~Lett.}}\
  }%
  \textbf{\bibinfo {volume} {96}},\ \bibinfo {pages} {115002--+} (\bibinfo
  {month} {Mar.}\ \bibinfo {year} {2006}),\
  \Eprint{http://arxiv.org/abs/arXiv:astro-ph/0511290}{arXiv:astro-ph/0511290}%
\BibitemShut{NoStop}%
\bibitem{Schekochihin:2004d}%
  \BibitemOpen
  \bibfield{author}{%
  \bibinfo {author} {\bibfnamefont{A.~A.}\ \bibnamefont{{Schekochihin}}},
  \bibinfo {author} {\bibfnamefont{S.~C.}\ \bibnamefont{{Cowley}}}, \bibinfo
  {author} {\bibfnamefont{S.~F.}\ \bibnamefont{{Taylor}}}, \bibinfo {author}
  {\bibfnamefont{J.~L.}\ \bibnamefont{{Maron}}},\ and\ \bibinfo {author}
  {\bibfnamefont{J.~C.}\ \bibnamefont{{McWilliams}}},\ }%
  \bibfield{title}{%
  \enquote{\bibinfo {title} {{Simulations of the Small-Scale Turbulent
  Dynamo}},}\ }%
  \bibfield{journal}{%
  \Doi{10.1086/422547}{\bibinfo {journal} {Astrophys.~J.}}\ }%
  \textbf{\bibinfo {volume} {612}},\ \bibinfo {pages} {276--307} (\bibinfo
  {month} {Sep.}\ \bibinfo {year} {2004}),\
  \Eprint{http://arxiv.org/abs/arXiv:astro-ph/0312046}{arXiv:astro-ph/0312046}%
\BibitemShut{NoStop}%
\bibitem{Note1}%
  \BibitemOpen
  \bibinfo {note} {It is important to note that the validity of this cascade
  model is \emph{not} limited to the validity of the gyrokinetic approximation,
  $k_\parallel \ll k_\perp $. At all inertial range scales $k \rho _i \ll 1$,
  the simplification $\omega = \pm \protect \overline {\omega }(k_\perp )
  k_\parallel v_A$ is true for the more general Vlasov-Maxwell system. If the
  Vlasov-Maxwell dispersion relation is used for the cascade model, the only
  limitation occurs when $\omega \rightarrow \Omega _i$ because the nature of
  the nonlinear energy transfer is unknown in this ion cyclotron
  regime.}\BibitemShut{Stop}%
\bibitem{Galtier:2006}%
  \BibitemOpen
  \bibfield{author}{%
  \bibinfo {author} {\bibfnamefont{S.}~\bibnamefont{{Galtier}}},\ }%
  \bibfield{title}{%
  \enquote{\bibinfo {title} {{Wave turbulence in incompressible Hall
  magnetohydrodynamics}},}\ }%
  \bibfield{journal}{%
  \Doi{10.1017/S0022377806004521}{\bibinfo {journal} {J.~Plasma Phys.}}\ }%
  \textbf{\bibinfo {volume} {72}},\ \bibinfo {pages} {721--769} (\bibinfo
  {year} {2006})\BibitemShut{NoStop}%
\bibitem{Gary:2008}%
  \BibitemOpen
  \bibfield{author}{%
  \bibinfo {author} {\bibfnamefont{S.~P.}\ \bibnamefont{{Gary}}}, \bibinfo
  {author} {\bibfnamefont{S.}~\bibnamefont{{Saito}}},\ and\ \bibinfo {author}
  {\bibfnamefont{H.}~\bibnamefont{{Li}}},\ }%
  \bibfield{title}{%
  \enquote{\bibinfo {title} {{Cascade of whistler turbulence: Particle-in-cell
  simulations}},}\ }%
  \bibfield{journal}{%
  \Doi{10.1029/2007GL032327}{\bibinfo {journal} {Geophys.~Res.~Lett.}}\ }%
  \textbf{\bibinfo {volume} {35}},\ \bibinfo {pages} {2104--+} (\bibinfo
  {month} {Jan.}\ \bibinfo {year} {2008})\BibitemShut{NoStop}%
\bibitem{Saito:2008}%
  \BibitemOpen
  \bibfield{author}{%
  \bibinfo {author} {\bibfnamefont{S.}~\bibnamefont{{Saito}}}, \bibinfo
  {author} {\bibfnamefont{S.~P.}\ \bibnamefont{{Gary}}}, \bibinfo {author}
  {\bibfnamefont{H.}~\bibnamefont{{Li}}},\ and\ \bibinfo {author}
  {\bibfnamefont{Y.}~\bibnamefont{{Narita}}},\ }%
  \bibfield{title}{%
  \enquote{\bibinfo {title} {{Whistler turbulence: Particle-in-cell
  simulations}},}\ }%
  \bibfield{journal}{%
  \Doi{10.1063/1.2997339}{\bibinfo {journal} {Phys.~Plasmas}}\ }%
  \textbf{\bibinfo {volume} {15}},\ \bibinfo {pages} {102305--+} (\bibinfo
  {month} {Oct.}\ \bibinfo {year} {2008})\BibitemShut{NoStop}%
\bibitem{Gary:2010}%
  \BibitemOpen
  \bibfield{author}{%
  \bibinfo {author} {\bibfnamefont{S.~P.}\ \bibnamefont{{Gary}}}, \bibinfo
  {author} {\bibfnamefont{S.}~\bibnamefont{{Saito}}},\ and\ \bibinfo {author}
  {\bibfnamefont{Y.}~\bibnamefont{{Narita}}},\ }%
  \bibfield{title}{%
  \enquote{\bibinfo {title} {{Whistler Turbulence Wavevector Anisotropies:
  Particle-in-cell Simulations}},}\ }%
  \bibfield{journal}{%
  \Doi{10.1088/0004-637X/716/2/1332}{\bibinfo {journal} {Astrophys.~J.}}\ }%
  \textbf{\bibinfo {volume} {716}},\ \bibinfo {pages} {1332--1335} (\bibinfo
  {month} {Jun.}\ \bibinfo {year} {2010})\BibitemShut{NoStop}%
\bibitem{Note6}%
  \BibitemOpen
  \bibinfo {note} {It is possible that strong collisionless damping can lead to
  a broadening of the three-wave resonances constraining weak turbulent
  interactions, and that this effect might lead to a parallel cascade of
  kinetic \Alfven waves in the weak turbulence limit. Nonlinear kinetic
  simulations focused on this problem will likely be able to settle this
  question. Until evidence is found in support of this possibility, however, we
  maintain the conjecture the weak kinetic \Alfven wave turbulence generates no
  parallel cascade.}\BibitemShut{Stop}%
\bibitem{Note2}%
  \BibitemOpen
  \bibinfo {note} {Note that coefficients $2/3$ and $1/3$ in this formula for
  $k_\parallel (k_\perp )$ are particular to the specified Goldreich-Sridhar
  model of anisotropic plasma turbulence \cite{Goldreich:1995}; for an
  alternative turbulence model, such as the Boldyrev theory
  \cite{Boldyrev:2006}, one needs merely to change these two coefficients
  appropriately.}\BibitemShut{Stop}%
\bibitem{Howes:2006}%
  \BibitemOpen
  \bibfield{author}{%
  \bibinfo {author} {\bibfnamefont{G.~G.}\ \bibnamefont{{Howes}}}, \bibinfo
  {author} {\bibfnamefont{S.~C.}\ \bibnamefont{{Cowley}}}, \bibinfo {author}
  {\bibfnamefont{W.}~\bibnamefont{{Dorland}}}, \bibinfo {author}
  {\bibfnamefont{G.~W.}\ \bibnamefont{{Hammett}}}, \bibinfo {author}
  {\bibfnamefont{E.}~\bibnamefont{{Quataert}}},\ and\ \bibinfo {author}
  {\bibfnamefont{A.~A.}\ \bibnamefont{{Schekochihin}}},\ }%
  \bibfield{title}{%
  \enquote{\bibinfo {title} {{Astrophysical Gyrokinetics: Basic Equations and
  Linear Theory}},}\ }%
  \bibfield{journal}{%
  \Doi{10.1086/506172}{\bibinfo {journal} {Astrophys.~J.}}\ }%
  \textbf{\bibinfo {volume} {651}},\ \bibinfo {pages} {590--614} (\bibinfo
  {month} {Nov.}\ \bibinfo {year} {2006}),\
  \Eprint{http://arxiv.org/abs/astro-ph/0511812}{astro-ph/0511812}\BibitemShut%
{NoStop}%
\bibitem{Numata:2010}%
  \BibitemOpen
  \bibfield{author}{%
  \bibinfo {author} {\bibfnamefont{R.}~\bibnamefont{{Numata}}}, \bibinfo
  {author} {\bibfnamefont{G.~G.}\ \bibnamefont{{Howes}}}, \bibinfo {author}
  {\bibfnamefont{T.}~\bibnamefont{{Tatsuno}}}, \bibinfo {author}
  {\bibfnamefont{M.}~\bibnamefont{{Barnes}}},\ and\ \bibinfo {author}
  {\bibfnamefont{W.}~\bibnamefont{{Dorland}}},\ }%
  \bibfield{title}{%
  \enquote{\bibinfo {title} {{AstroGK: Astrophysical gyrokinetics code}},}\ }%
  \bibfield{journal}{%
  \Doi{10.1016/j.jcp.2010.09.006}{\bibinfo {journal} {J.~Comp.~Phys.}}\ }%
  \textbf{\bibinfo {volume} {229}},\ \bibinfo {pages} {9347--9372} (\bibinfo
  {month} {Dec.}\ \bibinfo {year} {2010}),\
  \Eprint{http://arxiv.org/abs/1004.0279}{arXiv:1004.0279
  [physics.plasm-ph]}\BibitemShut{NoStop}%
\bibitem{Frieman:1982}%
  \BibitemOpen
  \bibfield{author}{%
  \bibinfo {author} {\bibfnamefont{E.~A.}\ \bibnamefont{{Frieman}}}\ and\
  \bibinfo {author} {\bibfnamefont{L.}~\bibnamefont{{Chen}}},\ }%
  \bibfield{title}{%
  \enquote{\bibinfo {title} {{Nonlinear gyrokinetic equations for low-frequency
  electromagnetic waves in general plasma equilibria}},}\ }%
  \bibfield{journal}{%
  \bibinfo {journal} {Phys.~Fluids}\ }%
  \textbf{\bibinfo {volume} {25}},\ \bibinfo {pages} {502--508} (\bibinfo
  {month} {Mar.}\ \bibinfo {year} {1982})\BibitemShut{NoStop}%
\bibitem{Abel:2008}%
  \BibitemOpen
  \bibfield{author}{%
  \bibinfo {author} {\bibfnamefont{I.~G.}\ \bibnamefont{{Abel}}}, \bibinfo
  {author} {\bibfnamefont{M.}~\bibnamefont{{Barnes}}}, \bibinfo {author}
  {\bibfnamefont{S.~C.}\ \bibnamefont{{Cowley}}}, \bibinfo {author}
  {\bibfnamefont{W.}~\bibnamefont{{Dorland}}},\ and\ \bibinfo {author}
  {\bibfnamefont{A.~A.}\ \bibnamefont{{Schekochihin}}},\ }%
  \bibfield{title}{%
  \enquote{\bibinfo {title} {{Linearized model Fokker-Planck collision
  operators for gyrokinetic simulations. I. Theory}},}\ }%
  \bibfield{journal}{%
  \Doi{10.1063/1.3046067}{\bibinfo {journal} {Phys.~Plasmas}}\ }%
  \textbf{\bibinfo {volume} {15}},\ \bibinfo {pages} {122509--+} (\bibinfo
  {month} {Dec.}\ \bibinfo {year} {2008}),\
  \Eprint{http://arxiv.org/abs/0808.1300}{arXiv:0808.1300}\BibitemShut{NoStop}%
\bibitem{Barnes:2009}%
  \BibitemOpen
  \bibfield{author}{%
  \bibinfo {author} {\bibfnamefont{M.}~\bibnamefont{{Barnes}}}, \bibinfo
  {author} {\bibfnamefont{I.~G.}\ \bibnamefont{{Abel}}}, \bibinfo {author}
  {\bibfnamefont{W.}~\bibnamefont{{Dorland}}}, \bibinfo {author}
  {\bibfnamefont{D.~R.}\ \bibnamefont{{Ernst}}}, \bibinfo {author}
  {\bibfnamefont{G.~W.}\ \bibnamefont{{Hammett}}}, \bibinfo {author}
  {\bibfnamefont{P.}~\bibnamefont{{Ricci}}}, \bibinfo {author}
  {\bibfnamefont{B.~N.}\ \bibnamefont{{Rogers}}}, \bibinfo {author}
  {\bibfnamefont{A.~A.}\ \bibnamefont{{Schekochihin}}},\ and\ \bibinfo {author}
  {\bibfnamefont{T.}~\bibnamefont{{Tatsuno}}},\ }%
  \bibfield{title}{%
  \enquote{\bibinfo {title} {{Linearized model Fokker-Planck collision
  operators for gyrokinetic simulations. II. Numerical implementation and
  tests}},}\ }%
  \bibfield{journal}{%
  \Doi{10.1063/1.3155085}{\bibinfo {journal} {Phys.~Plasmas}}\ }%
  \textbf{\bibinfo {volume} {16}},\ \bibinfo {pages} {072107--+} (\bibinfo
  {month} {Jul.}\ \bibinfo {year} {2009})\BibitemShut{NoStop}%
\bibitem{Note3}%
  \BibitemOpen
  \bibinfo {note} {Because the window defining the local contributions spans
  $[k_\perp /2,2k_\perp ]$, there is an unusual peak at the left for $\epsilon
  _l/\epsilon $ where the entire window first falls within the range of scales
  modeled; a similar peak can be seen in the undamped model at the right
  side.}\BibitemShut{Stop}%
\bibitem{Note4}%
  \BibitemOpen
  \bibinfo {note} {Note that nonlocal motions at large scales need only be a
  factor of two or more larger than the local scale, so even for our
  simulations with a modest perpendicular dynamic range of 42, nonlocal motions
  at large scales can significantly contribute to the energy transfer
  rate.}\BibitemShut{Stop}%
\bibitem{Alexakis:2005}%
  \BibitemOpen
  \bibfield{author}{%
  \bibinfo {author} {\bibfnamefont{A.}~\bibnamefont{{Alexakis}}}, \bibinfo
  {author} {\bibfnamefont{P.~D.}\ \bibnamefont{{Mininni}}},\ and\ \bibinfo
  {author} {\bibfnamefont{A.}~\bibnamefont{{Pouquet}}},\ }%
  \bibfield{title}{%
  \enquote{\bibinfo {title} {{Shell-to-shell energy transfer in
  magnetohydrodynamics. I. Steady state turbulence}},}\ }%
  \bibfield{journal}{%
  \Doi{10.1103/PhysRevE.72.046301}{\bibinfo {journal} {Phys.~Rev.~E}}\ }%
  \textbf{\bibinfo {volume} {72}},\ \bibinfo {pages} {046301--+} (\bibinfo
  {month} {Oct.}\ \bibinfo {year} {2005})\BibitemShut{NoStop}%
\bibitem{Carati:2006}%
  \BibitemOpen
  \bibfield{author}{%
  \bibinfo {author} {\bibfnamefont{D.}~\bibnamefont{{Carati}}}, \bibinfo
  {author} {\bibfnamefont{O.}~\bibnamefont{{Debliquy}}}, \bibinfo {author}
  {\bibfnamefont{B.}~\bibnamefont{{Knaepen}}}, \bibinfo {author}
  {\bibfnamefont{B.}~\bibnamefont{{Teaca}}},\ and\ \bibinfo {author}
  {\bibfnamefont{M.}~\bibnamefont{{Verma}}},\ }%
  \bibfield{title}{%
  \enquote{\bibinfo {title} {{Energy transfers in forced MHD turbulence}},}\ }%
  \bibfield{journal}{%
  \Doi{10.1080/14685240600774017}{\bibinfo {journal} {Journal of Turbulence}}\
  }%
  \textbf{\bibinfo {volume} {7}},\ \bibinfo {pages} {51--+} (\bibinfo {year}
  {2006})\BibitemShut{NoStop}%
\bibitem{Yousef:2007}%
  \BibitemOpen
  \bibfield{author}{%
  \bibinfo {author} {\bibfnamefont{T.~A.}\ \bibnamefont{{Yousef}}}, \bibinfo
  {author} {\bibfnamefont{F.}~\bibnamefont{{Rincon}}},\ and\ \bibinfo {author}
  {\bibfnamefont{A.~A.}\ \bibnamefont{{Schekochihin}}},\ }%
  \bibfield{title}{%
  \enquote{\bibinfo {title} {{Exact scaling laws and the local structure of
  isotropic magnetohydrodynamic turbulence}},}\ }%
  \bibfield{journal}{%
  \Doi{10.1017/S0022112006004186}{\bibinfo {journal} {J.~Fluid Mech.}}\ }%
  \textbf{\bibinfo {volume} {575}},\ \bibinfo {pages} {111--+} (\bibinfo
  {month} {Mar.}\ \bibinfo {year} {2007}),\
  \Eprint{http://arxiv.org/abs/arXiv:astro-ph/0611692}{arXiv:astro-ph/0611692}%
\BibitemShut{NoStop}%
\bibitem{Aluie:2010}%
  \BibitemOpen
  \bibfield{author}{%
  \bibinfo {author} {\bibfnamefont{H.}~\bibnamefont{{Aluie}}}\ and\ \bibinfo
  {author} {\bibfnamefont{G.~L.}\ \bibnamefont{{Eyink}}},\ }%
  \bibfield{title}{%
  \enquote{\bibinfo {title} {{Scale Locality of Magnetohydrodynamic
  Turbulence}},}\ }%
  \bibfield{journal}{%
  \Doi{10.1103/PhysRevLett.104.081101}{\bibinfo {journal} {Phys.~Rev.~Lett.}}\
  }%
  \textbf{\bibinfo {volume} {104}},\ \bibinfo {pages} {081101--+} (\bibinfo
  {month} {Feb.}\ \bibinfo {year} {2010}),\
  \Eprint{http://arxiv.org/abs/0912.3752}{arXiv:0912.3752}\BibitemShut{NoStop}%
\bibitem{Mininni:2010}%
  \BibitemOpen
  \bibfield{author}{%
  \bibinfo {author} {\bibfnamefont{P.~D.}\ \bibnamefont{{Mininni}}},\ }%
  \bibfield{title}{%
  \enquote{\bibinfo {title} {{Scale Interactions in Magnetohydrodynamic
  Turbulence}},}\ }%
  \bibfield{journal}{%
  \Doi{10.1146/annurev-fluid-122109-160748}{\bibinfo {journal} {Ann. Rev. of
  Fluid Mech.}}\ }%
  \textbf{\bibinfo {volume} {43}},\ \bibinfo {pages} {377--397} (\bibinfo
  {month} {Jan.}\ \bibinfo {year} {2011}),\
  \Eprint{http://arxiv.org/abs/1006.1817}{arXiv:1006.1817
  [physics.flu-dyn]}\BibitemShut{NoStop}%
\bibitem{Podesta:2010a}%
  \BibitemOpen
  \bibfield{author}{%
  \bibinfo {author} {\bibfnamefont{J.~J.}\ \bibnamefont{{Podesta}}}, \bibinfo
  {author} {\bibfnamefont{J.~E.}\ \bibnamefont{{Borovsky}}},\ and\ \bibinfo
  {author} {\bibfnamefont{S.~P.}\ \bibnamefont{{Gary}}},\ }%
  \bibfield{title}{%
  \enquote{\bibinfo {title} {{A Kinetic Alfv{\'e}n Wave Cascade Subject to
  Collisionless Damping Cannot Reach Electron Scales in the Solar Wind at 1
  AU}},}\ }%
  \bibfield{journal}{%
  \Doi{10.1088/0004-637X/712/1/685}{\bibinfo {journal} {Astrophys.~J.}}\ }%
  \textbf{\bibinfo {volume} {712}},\ \bibinfo {pages} {685--691} (\bibinfo
  {month} {Mar.}\ \bibinfo {year} {2010}),\
  \Eprint{http://arxiv.org/abs/0912.4026}{arXiv:0912.4026}\BibitemShut{NoStop}%
\bibitem{Note5}%
  \BibitemOpen
  \bibinfo {note} {We do not compare to the cascade rate based on their
  equation~(7) because their inclusion of the factor $\alpha ^2$ double counts
  the effect of dissipation on the cascade rate, which is already included by
  the decrease in amplitude of $E(k)$.}\BibitemShut{Stop}%
\bibitem{Mason:2006}%
  \BibitemOpen
  \bibfield{author}{%
  \bibinfo {author} {\bibfnamefont{J.}~\bibnamefont{{Mason}}}, \bibinfo
  {author} {\bibfnamefont{F.}~\bibnamefont{{Cattaneo}}},\ and\ \bibinfo
  {author} {\bibfnamefont{S.}~\bibnamefont{{Boldyrev}}},\ }%
  \bibfield{title}{%
  \enquote{\bibinfo {title} {{Dynamic Alignment in Driven Magnetohydrodynamic
  Turbulence}},}\ }%
  \bibfield{journal}{%
  \Doi{10.1103/PhysRevLett.97.255002}{\bibinfo {journal} {Phys.~Rev.~Lett.}}\
  }%
  \textbf{\bibinfo {volume} {97}},\ \bibinfo {pages} {255002--+} (\bibinfo
  {month} {Dec.}\ \bibinfo {year} {2006}),\
  \Eprint{http://arxiv.org/abs/arXiv:astro-ph/0602382}{arXiv:astro-ph/0602382}%
\BibitemShut{NoStop}%
\bibitem{Mason:2008}%
  \BibitemOpen
  \bibfield{author}{%
  \bibinfo {author} {\bibfnamefont{J.}~\bibnamefont{{Mason}}}, \bibinfo
  {author} {\bibfnamefont{F.}~\bibnamefont{{Cattaneo}}},\ and\ \bibinfo
  {author} {\bibfnamefont{S.}~\bibnamefont{{Boldyrev}}},\ }%
  \bibfield{title}{%
  \enquote{\bibinfo {title} {{Numerical measurements of the spectrum in
  magnetohydrodynamic turbulence}},}\ }%
  \bibfield{journal}{%
  \Doi{10.1103/PhysRevE.77.036403}{\bibinfo {journal} {Phys.~Rev.~E}}\ }%
  \textbf{\bibinfo {volume} {77}},\ \bibinfo {pages} {036403--+} (\bibinfo
  {month} {Mar.}\ \bibinfo {year} {2008}),\
  \Eprint{http://arxiv.org/abs/0706.2003}{arXiv:0706.2003}\BibitemShut{NoStop}%
\bibitem{Boldyrev:2009}%
  \BibitemOpen
  \bibfield{author}{%
  \bibinfo {author} {\bibfnamefont{S.}~\bibnamefont{{Boldyrev}}}, \bibinfo
  {author} {\bibfnamefont{J.}~\bibnamefont{{Mason}}},\ and\ \bibinfo {author}
  {\bibfnamefont{F.}~\bibnamefont{{Cattaneo}}},\ }%
  \bibfield{title}{%
  \enquote{\bibinfo {title} {{Dynamic Alignment and Exact Scaling Laws in
  Magnetohydrodynamic Turbulence}},}\ }%
  \bibfield{journal}{%
  \Doi{10.1088/0004-637X/699/1/L39}{\bibinfo {journal} {Astrophys.~J.~Lett.}}\
  }%
  \textbf{\bibinfo {volume} {699}},\ \bibinfo {pages} {L39--L42} (\bibinfo
  {month} {Jul.}\ \bibinfo {year} {2009})\BibitemShut{NoStop}%
\bibitem{Perez:2009}%
  \BibitemOpen
  \bibfield{author}{%
  \bibinfo {author} {\bibfnamefont{J.~C.}\ \bibnamefont{{Perez}}}\ and\
  \bibinfo {author} {\bibfnamefont{S.}~\bibnamefont{{Boldyrev}}},\ }%
  \bibfield{title}{%
  \enquote{\bibinfo {title} {{Role of Cross-Helicity in Magnetohydrodynamic
  Turbulence}},}\ }%
  \bibfield{journal}{%
  \Doi{10.1103/PhysRevLett.102.025003}{\bibinfo {journal} {Phys.~Rev.~Lett.}}\
  }%
  \textbf{\bibinfo {volume} {102}},\ \bibinfo {pages} {025003--+} (\bibinfo
  {month} {Jan.}\ \bibinfo {year} {2009}),\
  \Eprint{http://arxiv.org/abs/0807.2635}{arXiv:0807.2635}\BibitemShut{NoStop}%
\bibitem{Tatsuno:2009}%
  \BibitemOpen
  \bibfield{author}{%
  \bibinfo {author} {\bibfnamefont{T.}~\bibnamefont{{Tatsuno}}}, \bibinfo
  {author} {\bibfnamefont{A.~A.}\ \bibnamefont{{Schekochihin}}}, \bibinfo
  {author} {\bibfnamefont{W.}~\bibnamefont{{Dorland}}}, \bibinfo {author}
  {\bibfnamefont{G.}~\bibnamefont{{Plunk}}}, \bibinfo {author}
  {\bibfnamefont{M.~A.}\ \bibnamefont{{Barnes}}}, \bibinfo {author}
  {\bibfnamefont{S.~C.}\ \bibnamefont{{Cowley}}},\ and\ \bibinfo {author}
  {\bibfnamefont{G.~G.}\ \bibnamefont{{Howes}}},\ }%
  \bibfield{title}{%
  \enquote{\bibinfo {title} {{Nonlinear phase mixing and phase-space cascade of
  entropy in gyrokinetic plasma turbulence}},}\ }%
  \bibfield{journal}{%
  \Doi{10.1103/PhysRevLett.103.015003}{\bibinfo {journal} {Phys.~Rev.~Lett.}}\
  }%
  \textbf{\bibinfo {volume} {103}},\ \bibinfo {pages} {015003} (\bibinfo {year}
  {2009})\BibitemShut{NoStop}%
\bibitem{Plunk:2010}%
  \BibitemOpen
  \bibfield{author}{%
  \bibinfo {author} {\bibfnamefont{G.~G.}\ \bibnamefont{{Plunk}}}, \bibinfo
  {author} {\bibfnamefont{S.~C.}\ \bibnamefont{{Cowley}}}, \bibinfo {author}
  {\bibfnamefont{A.~A.}\ \bibnamefont{{Schekochihin}}},\ and\ \bibinfo {author}
  {\bibfnamefont{T.}~\bibnamefont{{Tatsuno}}},\ }%
  \bibfield{title}{%
  \enquote{\bibinfo {title} {{Two-dimensional gyrokinetic turbulence}},}\ }%
  \bibfield{journal}{%
  \Doi{10.1017/S002211201000371X}{\bibinfo {journal} {Journal of Fluid
  Mechanics}}\ }%
  \textbf{\bibinfo {volume} {664}},\ \bibinfo {pages} {407--435} (\bibinfo
  {month} {Dec.}\ \bibinfo {year} {2010}),\
  \Eprint{http://arxiv.org/abs/0904.0243}{arXiv:0904.0243
  [physics.plasm-ph]}\BibitemShut{NoStop}%
\bibitem{Tatsuno:2010}%
  \BibitemOpen
  \bibfield{author}{%
  \bibinfo {author} {\bibfnamefont{T.}~\bibnamefont{{Tatsuno}}}, \bibinfo
  {author} {\bibfnamefont{M.}~\bibnamefont{{Barnes}}}, \bibinfo {author}
  {\bibfnamefont{S.~C.}\ \bibnamefont{{Cowley}}}, \bibinfo {author}
  {\bibfnamefont{W.}~\bibnamefont{{Dorland}}}, \bibinfo {author}
  {\bibfnamefont{G.~G.}\ \bibnamefont{{Howes}}}, \bibinfo {author}
  {\bibfnamefont{R.}~\bibnamefont{{Numata}}}, \bibinfo {author}
  {\bibfnamefont{G.~G.}\ \bibnamefont{{Plunk}}},\ and\ \bibinfo {author}
  {\bibfnamefont{A.~A.}\ \bibnamefont{{Schekochihin}}},\ }%
  \bibfield{title}{%
  \enquote{\bibinfo {title} {{Gyrokinetic simulation of entropy cascade in
  two-dimensional electrostatic turbulence}},}\ }%
  \bibfield{journal}{%
  \bibinfo {journal} {J. Plasma Fusion Res.}}%
   (\bibinfo {month} {Mar.}\ \bibinfo {year} {2010}),\ \bibinfo {note}
  {accepted},\
  \Eprint{http://arxiv.org/abs/1003.3933}{arXiv:1003.3933}\BibitemShut{NoStop}%
\bibitem{Plunk:2011}%
  \BibitemOpen
  \bibfield{author}{%
  \bibinfo {author} {\bibfnamefont{G.~G.}\ \bibnamefont{{Plunk}}}\ and\
  \bibinfo {author} {\bibfnamefont{T.}~\bibnamefont{{Tatsuno}}},\ }%
  \bibfield{title}{%
  \enquote{\bibinfo {title} {{Energy Transfer and Dual Cascade in Kinetic
  Magnetized Plasma Turbulence}},}\ }%
  \bibfield{journal}{%
  \Doi{10.1103/PhysRevLett.106.165003}{\bibinfo {journal} {Phys.~Rev.~Lett.}}\
  }%
  \textbf{\bibinfo {volume} {106}},\ \bibinfo {pages} {165003--+} (\bibinfo
  {month} {Apr.}\ \bibinfo {year} {2011}),\
  \Eprint{http://arxiv.org/abs/1007.4787}{arXiv:1007.4787
  [physics.plasm-ph]}\BibitemShut{NoStop}%
\bibitem{Bale:2005}%
  \BibitemOpen
  \bibfield{author}{%
  \bibinfo {author} {\bibfnamefont{S.~D.}\ \bibnamefont{{Bale}}}, \bibinfo
  {author} {\bibfnamefont{P.~J.}\ \bibnamefont{{Kellogg}}}, \bibinfo {author}
  {\bibfnamefont{F.~S.}\ \bibnamefont{{Mozer}}}, \bibinfo {author}
  {\bibfnamefont{T.~S.}\ \bibnamefont{{Horbury}}},\ and\ \bibinfo {author}
  {\bibfnamefont{H.}~\bibnamefont{{Reme}}},\ }%
  \bibfield{title}{%
  \enquote{\bibinfo {title} {{Measurement of the Electric Fluctuation Spectrum
  of Magnetohydrodynamic Turbulence}},}\ }%
  \bibfield{journal}{%
  \Doi{10.1103/PhysRevLett.94.215002}{\bibinfo {journal} {Phys.~Rev.~Lett.}}\
  }%
  \textbf{\bibinfo {volume} {94}},\ \bibinfo {pages} {215002--+} (\bibinfo
  {month} {Jun.}\ \bibinfo {year} {2005}),\
  \Eprint{http://arxiv.org/abs/physics/0503103}{physics/0503103}\BibitemShut{N%
oStop}%
\bibitem{Stawicki:2001}%
  \BibitemOpen
  \bibfield{author}{%
  \bibinfo {author} {\bibfnamefont{O.}~\bibnamefont{{Stawicki}}}, \bibinfo
  {author} {\bibfnamefont{S.~P.}\ \bibnamefont{{Gary}}},\ and\ \bibinfo
  {author} {\bibfnamefont{H.}~\bibnamefont{{Li}}},\ }%
  \bibfield{title}{%
  \enquote{\bibinfo {title} {{Solar wind magnetic fluctuation spectra:
  Dispersion versus damping}},}\ }%
  \bibfield{journal}{%
  \Doi{10.1029/2000JA000446}{\bibinfo {journal} {J.~Geophys.~Res.}}\ }%
  \textbf{\bibinfo {volume} {106}},\ \bibinfo {pages} {8273--8282} (\bibinfo
  {month} {May}\ \bibinfo {year} {2001})\BibitemShut{NoStop}%
\bibitem{Rudakov:2011}%
  \BibitemOpen
  \bibfield{author}{%
  \bibinfo {author} {\bibfnamefont{L.}~\bibnamefont{{Rudakov}}}, \bibinfo
  {author} {\bibfnamefont{M.}~\bibnamefont{{Mithaiwala}}}, \bibinfo {author}
  {\bibfnamefont{G.}~\bibnamefont{{Ganguli}}},\ and\ \bibinfo {author}
  {\bibfnamefont{C.}~\bibnamefont{{Crabtree}}},\ }%
  \bibfield{title}{%
  \enquote{\bibinfo {title} {{Linear and nonlinear Landau resonance of kinetic
  Alfv{\'e}n waves: Consequences for electron distribution and wave spectrum in
  the solar wind}},}\ }%
  \bibfield{journal}{%
  \Doi{10.1063/1.3532819}{\bibinfo {journal} {Phys.~Plasmas}}\ }%
  \textbf{\bibinfo {volume} {18}},\ \bibinfo {pages} {012307--+} (\bibinfo
  {month} {Jan.}\ \bibinfo {year} {2011}),\
  \Eprint{http://arxiv.org/abs/1008.0993}{arXiv:1008.0993
  [astro-ph.SR]}\BibitemShut{NoStop}%
\end{thebibliography}
%

\end{document}